\documentclass[pra, twocolumn, showpacs, amsmath, amssymb , noshowkeys]{revtex4}

\usepackage{dcolumn}
\usepackage{bm}
\RequirePackage[usenames,dvipsnames]{color}

\usepackage{color}
\usepackage{epsfig} 
\psfull                                
\usepackage{mathbbol}
\usepackage{bbm}
\usepackage{dsfont}

\newcommand{\bra}[1]{\langle #1|}	
\newcommand{\ket}[1]{|#1\rangle}
\newcommand{\braket}[2]{\langle #1|#2\rangle}

\newcommand{\rme}{\text{e}}
\newcommand{\rmi}{\text{i}}
\newcommand{\sign}[1]{\text{sign}\left(#1 \right)}
\newcommand{\fh}{\mathcal{H}}

\newcommand{\brafl}[1]{\langle \langle #1|}	
\newcommand{\ketfl}[1]{|#1\rangle \rangle}
\newcommand{\braketfl}[2]{\langle \langle #1|#2\rangle \rangle}

\newcommand{\eps}{\varepsilon}
\newcommand{\uMin}[1]{\ketfl{u_{\uparrow,#1}^0}}
\newcommand{\uPlus}[1]{\ketfl{u_{\downarrow,#1}^0}}
\newcommand{\ulfMin}[1]{\brafl{u_{\uparrow,#1}^0}}
\newcommand{\ulfPlus}[1]{\brafl{u_{\downarrow,#1}^0}}
\newcommand{\epsMin}[1]{\varepsilon^0_{\uparrow, #1}}
\newcommand{\epsPlus}[1]{\varepsilon^0_{\downarrow, #1}}
\newcommand{\DDelta}[1]{\Delta_{#1}}
\newcommand{\hh}{\frac{\hbar}{2}}

\newcommand{\X}[2]{X_{#1, #2}}
\newcommand{\PflMin}[2]{\ketfl{\Phi_{-,#1}^{#2}}}
\newcommand{\PflPlus}[2]{\ketfl{\Phi_{+,#1}^{#2}}}

\newcommand{\str}[1]{S^{(#1)}}
\newcommand{\ketT}[1]{|#1)}

\newcommand{\braketT}[2]{(#1|#2)}
\newcommand{\Mixangle}{\Theta_{m}}
\newcommand{\Mixang}[1]{\Theta_{m}^{#1}}

\newcommand{\sinmix}[1]{\sin \frac{\Mixang{#1}}{2}}
\newcommand{\cosmix}[1]{\cos \frac{\Mixang{#1}}{2}}
\newcommand{\sinmixa}{\sin \frac{\Mixangle}{2}}
\newcommand{\cosmixa}{\cos \frac{\Mixangle}{2}}
\newcommand{\half}{\frac{1}{2}}

\newcommand{\Dm}{\DDelta{-m}}

\newcommand{\LDiss}[2]{\mathcal{L}_{#1, #2}}
\newcommand{\LMRWA}[2]{\mathcal{L}_{#1, #2}^{\rm MRWA}}

\begin{document}

\title{The dissipative two-level system under strong ac-driving: a combination of Floquet and Van Vleck perturbation theory}
\date{\today}
\author{Johannes Hausinger} 
\email{johannes.hausinger@physik.uni-r.de}
\author{Milena Grifoni}
\affiliation{Institut f\"{u}r Theoretische Physik, Universit\"at
Regensburg, DE-93040 Regensburg, Germany}

\begin{abstract}
  We study the dissipative dynamics of a two-level system (TLS) exposed to strong ac driving. By combing Floquet theory with Van Vleck perturbation theory in the TLS tunneling matrix element, we diagonalize the time-dependent Hamiltonian  and  provide corrections to the renormalized Rabi frequency of the TLS, which are  valid for both a biased and unbiased TLS and go beyond the known high-frequency  and rotating-wave results.  In order to mimic environmental influences on the TLS, we couple the system weakly to a thermal bath and solve analytically the corresponding Floquet-Bloch-Redfield master equation. We give a closed expression for the relaxation and dephasing rates of the TLS and discuss their behavior under variation of the driving amplitude. Further, we examine the robustness of coherent destruction of tunneling (CDT) and driving-induced tunneling oscillations (DITO).  We show that also for a moderate driving frequency  an almost complete suppression of tunneling can be achieved for short times and demonstrate the sensitiveness of DITO to a change of the external parameters.  
\end{abstract}
\pacs{03.65.Yz, 42.50.Hz, 03.67.Lx}
\keywords{}
\maketitle

\section{Introduction}
The dissipative two-level system (TLS) has quite a rich and long history of both experimental and theoretical investigations \cite{Weiss2008}. Despite its simplicity, it is a very prominent candidate for modeling various different situations in physics as well as in chemistry and provides a testing ground for exploring dissipation and decoherence effects in genuine quantum-mechanical systems. The development of maser and laser technology triggered the examination of those systems under the influence of strong time-dependent driving fields, which yields dressed TLS states \cite{CohenTannoudji2004} in turn leading to a variety of phenomena like coherent destruction of tunneling (CDT) \cite{Grossmann1991(2),Grossmann1991(1),Grossmann1992} or driving-induced tunneling oscillations (DITO) \cite{Hartmann1998, Hartmann2000, Goychuk2005, Nakamura2001}. For taking into account the influence of the environment,  the driven spin-boson model \cite{Grifoni1998,Goychuk2005} has proven to be a suitable candidate.\newline
In recent years, the driven TLS has experienced a strong revival in the field of quantum computation, as lithographic fabrication techniques allow the construction of artificial atoms that are coupled to the modes of an oscillating field by a strength never reached in real atoms. Here the TLS implements the two logical states of a qubit. We mention just  two prominent solid-state realizations of the qubit, namely, the Cooper-pair box \cite{Nakamura1999, Makhlin2001, Vion2002, Collin2004} and the Josephson flux qubit \cite{Mooij1999, vanderWal2000, Chiorescu2003}. Strong coupling between the TLS and a single oscillator photon was first successfully reported in \cite{Wallraff2004} for a charge qubit.  In this weak driving limit, the oscillator is usually described in its quantized version and occupied by a small number of photons \cite{Blais2004}. Recently, a lot of theoretical effort has been put into solving the dynamics of such a system \cite{Tian2002, vanderWal2003, Kleff2003, Kleff2004, Wilhelm2004, Nesi2007, Brito2008, Hausinger2008, Huang2008}. Also for strong driving, the applied field can still be described by a quantized oscillator. However, for high photon numbers, the TLS-oscillator system is conveniently treated in the dressed state picture \cite{CohenTannoudji2004}. In this strong driving regime, a series of experiments and theoretical investigations have been performed recently on superconducting qubits examining Rabi oscillations in the multiphoton regime and the validity of the dressed state picture \cite{Nakamura2001,Saito2004, Saito2006, Izmalkov2004, Oliver2005, Berns2006, Sillanpaa2006, Wilson2007, Berns2008, Rudner2008, Wen2009, Wilson2009, Baur2009}.
To account for environmental effects, the latter is usually combined with the phenomenological Bloch equations \cite{Nakamura2001, Saito2004, Wilson2007, Wilson2009}. \newline 
With the applied field being in a coherent state and for high photon numbers, an equivalent description consists in replacing the quantized oscillator by an external, classical  driving \cite{Shirley1965, CohenTannoudji2004}. Together with the coupling to a bath of harmonic oscillators, it leads to the driven spin-boson model \cite{Grifoni1998, Goychuk2005}, which has been examined by applying various techniques. For example, the (real-time) path-integral formalism provides a formal, exact generalized master equation  for the dynamics of the reduced density matrix of the TLS, which can be solved approximately for certain parameter regimes \cite{Grifoni1993,  Grifoni1995, Makarov1995, Grifoni1996, Winterstetter1997, Grifoni1997, Grifoni1997(2), Hartmann1998, Hartmann2000}. Among those  treatments, the  noninteracting blip approximation (NIBA) \cite{Leggett1987, Weiss2008} is the most prominent one and is based on an expansion to lowest order in the tunneling matrix element of the undisturbed TLS. It provides good approximate results for intermediate to high bath temperatures and/or strong damping of the system with arbitrary driving frequencies. However, at low temperature  it fails to reproduce the dynamics of a biased TLS correctly.  In \cite{Dakhnovskii1994, Dakhnovskii1994(2), Dakhnovskii1994(3), Dakhnovskii1995, Wang1998}, the polaron transformation leads to an integro-differential kinetic equation for the populations of the density matrix, which is equivalent to the generalized master equation under the NIBA. An alternative way to gain the dynamics of the driven spin-boson model for weak system-bath coupling and within the Markovian limit is to solve the underlying Bloch-Redfield equations. This is done numerically for weak damping in \cite{Hartmann2000, Thorwart2000, Goorden2003}, while \cite{Hartmann2000, Goorden2003} additionally provide an analytical examination of the dynamics in the high-frequency regime.\newline
In this work we introduce a new approach to solving the dynamics of the monochromatically driven spin-boson model taking into account analytically the fast oscillations induced by the driving as well as the transient dynamics. In a first step, we combine Floquet theory \cite{Shirley1965, Sambe1973, Grifoni1998} with Van Vleck perturbation theory \cite{VanVleck1929, CohenTannoudji2004} to derive the dynamics of the nondissipative system. This approach has recently been used also in \cite{Son2009} to evaluate the time-averaged transition probability of a nondissipative TLS.   Going to second-order in the tunneling matrix element, we derive expressions which include the fast oscillatory behavior of the Floquet states and are beyond the common rotating-wave results  \cite{Ashhab2007, Oliver2005} or perturbation theory in the driving strength \cite{Aravind1984, Shirley1965}. Further, to analyze dissipative effects, we consider the regime of weak damping and solve the corresponding Floquet-Bloch-Redfield master equation applying a moderate rotating-wave approximation. While in \cite{Goorden2004, Goorden2005} a similar approach is used  to study the asymptotic dynamics of the driven spin-boson model perturbatively in the driving strength,  our approach treats the full time evolution of the system, to all orders in the driving amplitude, in the regime of moderate as well as high external frequencies and for arbitrary static bias. Specifically, we are able to give closed analytic expressions for both the relaxation and dephasing rates. \newline
Our analysis enables us to shed light on the famous effects of CDT \cite{Grossmann1991(2),Grossmann1991(1),Grossmann1992} and DITO \cite{Hartmann1998, Hartmann2000, Goychuk2005}. Many investigations of those phenomena have been performed in the high-driving regime. This work treats them  analytically also for moderate driving frequency and amplitude. We examine both the nondissipative and dissipative cases and  compare them to  a numerical solution of the problem.\newline
The structure of the work is a follows. In Sec. \ref{sec::NonDiss} the model Hamiltonian for the nondissipative system is introduced. We derive the corresponding Floquet Hamiltonian in Sec. \ref{sec::FloquetHamiltonian} and analyze its quasienergy spectrum and the dynamics of the system in Sec. \ref{sec::RWA} using a rotating-wave approximation (RWA). In Sec. \ref{sec::VanVleck} we apply Van Vleck perturbation theory to second-order in the tunneling matrix element  and compare the improved quasienergy spectrum to a numerical analysis.  Further, we give in Sec. \ref{sec::ValidVanVleck} a detailed discussion of the parameter regime in which our approach is valid. To tackle the dissipative dynamics, we introduce the driven spin-boson Hamiltonian in Sec. \ref{sec::DissSys} and solve the  Floquet-Bloch-Redfield equation. We compare the analytical expressions for the relaxation and dephasing rates to the results obtained within the RWA and close the paragraph with a discussion of CDT and DITO.\newline

\section{The nondissipative system} \label{sec::NonDiss}

In a first step, we neglect environmental effects on the driven TLS and consider the Hamiltonian 
 \begin{equation} \label{FreeHamil}
   H_{\text {TLS}}(t) = - \hh \left[ \Delta \sigma_x + (\varepsilon + A \cos \omega t) \sigma_z  \right].
\end{equation} 
Here, $\sigma_z$ and $\sigma_x$ are the  Pauli matrices, and as basis states we choose the eigenstates of $\sigma_z$, $\ket{\uparrow}$ and $\ket{\downarrow}$ (localized basis). The coupling strength $\Delta$ between those two basis states is time independent, whereas the bias point consists of the dc component $\eps$ and  a sinusoidal modulation of the amplitude $A$ and frequency $\omega$.

\subsection{Floquet Hamiltonian} \label{sec::FloquetHamiltonian}
To resolve the dynamics of the driven system, we take advantage of its periodicity and apply Floquet theory \cite{Shirley1965, Sambe1973, Grifoni1998}, about which we give a short overview in Appendix \ref{app::FloquetTheory}. For the driven TLS it leads to the Floquet Hamiltonian $
   \fh_{\text {TLS}}(t) = H_\text{TLS}(t) - \rmi \hbar \partial_t$.
Considering the case $\Delta = 0$, we find the following set of eigenstates of $\fh_{\text {TLS}}(t)$:
\begin{align} 
  \ket{u^0_{\uparrow/\downarrow,n}(t)} &= \ket{\uparrow / \downarrow} \exp \left[\pm \rmi \frac{A}{2 \omega} \sin \omega t - \rmi n \omega t \right] \nonumber \\
        &= \ket{\uparrow / \downarrow} \sum_k \rme^{\pm  \rmi k \omega t} J_{k} \left(\frac{A}{2 \omega } \right) \rme^{-\rmi n \omega t} \label{BasisU}
\end{align} 
with quasienergies $\hbar \varepsilon_{\uparrow/\downarrow, n}^0 = \mp \frac{\hbar}{2} \varepsilon - \hbar n \omega$. Here, $J_k(x)$ is the $k$th-order Bessel function.
In the composite Hilbert space $\fh \otimes \mathcal{T}$ \cite{Sambe1973}, which is introduced in Appendix \ref{app::FloquetTheory}, those states become 
\begin{equation} 
   \ketfl{u^0_{\uparrow/\downarrow,n}}  = \ket{\uparrow / \downarrow} \sum_l J_{\pm (n-l)} \left( \frac{A}{2 \omega} \right) \ketT{l},\label{BasisUHT}
\end{equation}
with the state vectors $\{ \ketT{l} \}$ being a basis for $\mathcal{T}$ and $\ketfl{\alpha_l} = \ket{\alpha} \otimes \ketT{l}$.
  For the case of a finite tunneling matrix element $\Delta$, the Floquet Hamiltonian $\fh_{\text{TLS}}$ is nondiagonal in the above basis (\ref{BasisUHT}) and becomes in matrix representation
\begin{widetext}
 \begin{eqnarray} \label{FloquetMatrix}
\fh_{\text{TLS}} = \hbar \left(\begin{array}{ c|cccccc|c}
                                   			\ddots & \uMin{n}  & \uPlus{n}  & \uMin{n+1} & \uPlus{n+1} & \uMin{n+2} & \uPlus{n+2} & \\ \hline
							& & & & & & \\
							\uMin{n} & \epsMin{n} & -\half \DDelta{0} & 0 & -\half \DDelta{-1} & 0 & - \half \DDelta{-2}  \\
							& & & & & & \\
							\uPlus{n} & - \half \DDelta{0} & \epsPlus{n} & - \half \DDelta{1} & 0 & -\half \DDelta{2} & 0  \\
							& & & & & & \\
							\uMin{n+1} & 0 & -\half \DDelta{1} & \epsMin{n+1} & -\half \DDelta{0} & 0 & -\half \DDelta{-1}  \\
							& & & & & & \\
							\uPlus{n+1} & - \half \DDelta{-1} & 0 & -\half \DDelta{0} & \epsPlus{n+1} & - \half \DDelta{1} & 0  \\
							& & & & & & \\
							\uMin{n+2} & 0 & -\half \DDelta{2} & 0 & - \half \DDelta{1} & \epsMin{n+2} & -\half \DDelta{0}   \\
							& & & & & & \\
							\uPlus{n+2} & - \half \DDelta{-2} & 0 & -\half \DDelta{-1} & 0 & -\half \DDelta{0} & \epsPlus{n+2}   \\
							& & & & & & \\ \hline
							& & & & & & & \ddots
                                                        \end{array} \right). \nonumber\\
\end{eqnarray}
\end{widetext}
We defined 
\begin{equation} \label{DressedDelta}
  \DDelta{n-l} \equiv \Delta \brafl{u_{\uparrow,n}^0}  \sigma_x \ketfl{u_{\downarrow,l}^0} =  J_{n-l} \left( \frac{A}{\omega} \right) \Delta,
\end{equation} 
where we used the relation \cite{Abram1972}
\begin{equation}
 J_{n} (u\pm v) = \sum_{k= -\infty}^\infty J_{n\mp k} (u) J_k (v).
\end{equation}
 To find the dynamics of the system, we have to diagonalize the Floquet matrix. In the remaining subsections we discuss two approximation schemes. A rotating-wave approximation scheme is discussed in Sec. \ref{sec::RWA}, while in Sec.  \ref{sec::VanVleck} Van Vleck perturbation theory is presented. We also show that the RWA results can be obtained with Van Vleck perturbation theory to lowest order in $\Delta$.

\subsection{Rotating-wave approximation} \label{sec::RWA}
 Let us look at the spectrum of the unperturbed problem ($\Delta = 0$). We notice that whenever the static bias fulfills the condition $\eps = m \omega$, the states $\uMin{n}$ and $\uPlus{n+m}$ are degenerate, as then
\begin{equation} \label{ResCond}
  \epsPlus{n+m} - \epsMin{n} =  \eps -  m \omega = 0.
\end{equation} 
In this case, we speak of an $m$-photon resonance.
As long as $\Delta$ is only a small perturbation, $\omega \gg \Delta$,  then $\fh_{\text{TLS}}$ will exhibit a similar energy spectrum. The main  corrections to the unperturbed Hamiltonian come from matrix elements connecting the (almost) degenerate levels. Thus, as a first approximation, we diagonalize an effective Hamiltonian, which consists of 2 $\times$ 2 blocks of the kind
\begin{equation} \label{Heff}
  \left( \begin{array}{c c}
    \hbar \epsMin{n} & -\hh \DDelta{-m} \\
  -\hh \DDelta{-m} & \hbar \epsPlus{n+m}
\end{array}
  \right),
\end{equation} 
and describes the energy states being connected by an $m$-photon resonance.\newline
  This result is  also obtained within the RWA scheme as introduced in \cite{Ashhab2007, Oliver2005}. In those works, the \textit{time-dependent} system Hamiltonian (\ref{FreeHamil}) is transformed to a rotating frame, and only terms fulfilling the resonance condition (\ref{ResCond}) are kept, while the fast-rotating components are neglected.  This RWA is different from the conventional  \textit{Rabi rotating-wave approximation,} which is  perturbative in the driving amplitude $A$, see, e.g., \cite{CohenTannoudji2004, Aravind1984}, and becomes exact for circularly polarized radiation. In contrast, the RWA we are using treats the driving amplitude nonperturbatively.\newline
\begin{figure}
 \includegraphics[height=.45\textwidth,angle=-90]{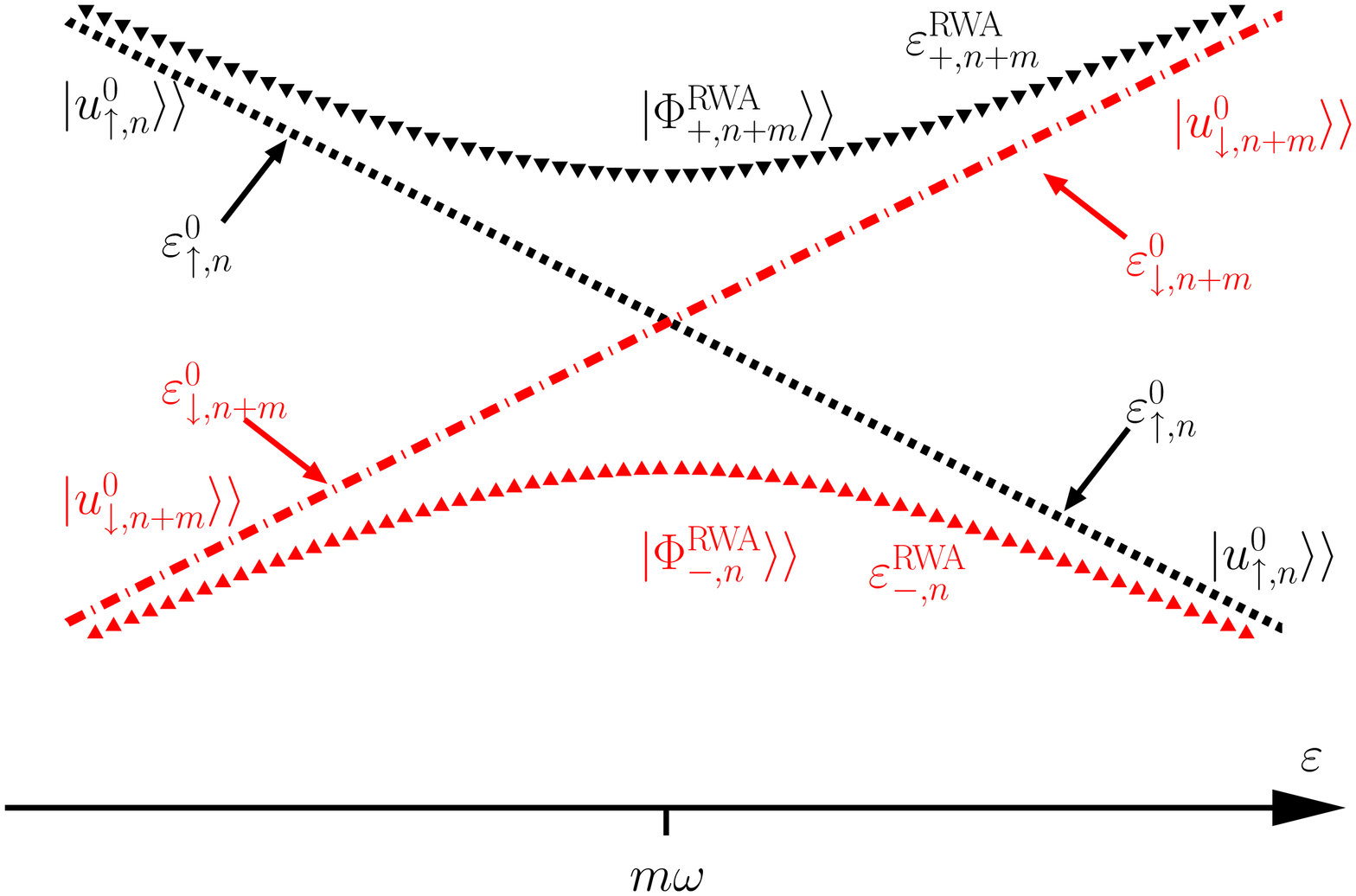}
 \caption{(Color online) Quasienergies $\eps_{-,n}^\text{RWA}$ and $\eps_{+,n+m}^\text{RWA}$ (triangles) and unperturbed quasienergies $\epsMin{n}$ and $\epsPlus{n+m}$ (dashed line and dotted-dashed line) for an $m$-photon resonance. The unperturbed quasienergies  show an exact crossing at $\eps = m \omega$ according to Eq. (\ref{ResCond}). The corresponding eigenstates are $\uMin{n}$ and $\uPlus{n+m}$. For finite $\Delta$ an avoided crossing can be observed.  The energy $\eps_{-,n}^\text{RWA}$ and the corresponding eigenstate $\ketfl{\Phi_{-,n}^\text{RWA}}$ are represented by black downward triangles, whereas $\eps_{+,n+m}^\text{RWA}$ and $\ketfl{\Phi_{+,n+m}^\text{RWA}}$ are shown by red upward triangles. For $\eps > m \omega$ we find  that $\ketfl{\Phi_{-,n}^\text{RWA}}$ approaches $\uMin{n}$, while $\ketfl{\Phi_{+,n+m}^\text{RWA}}$ becomes $\uPlus{n+m}$ and vice versa for $\eps < m \omega$. The labeling of the perturbed eigenstates and eigenenergies is chosen in a way that $\eps_{+,n+m}^\text{RWA} \geq \eps_{-,n}^\text{RWA}$ for all $\eps$. \label{fig::Avoided_Crossing}}
\end{figure}
Concerning the eigenenergies of the Floquet Hamiltonian for finite $\Delta$, we notice that the exact crossing of the unperturbed energies ($\Delta=0$) at $\eps = m \omega$ becomes an avoided crossing (see Fig. \ref{fig::Avoided_Crossing}) and the perturbed eigenstates are a mixture of the unperturbed ones. Those with higher eigenenergies are labeled $\ketfl{\Phi_{+,n+m}^\text{RWA}}$; those with lower energies $\ketfl{\Phi_{-,n}^\text{RWA}}$ \footnote{The motivation to choose the indices $n$ and $n+m$ of the perturbed eigenstates in this way is that they agree with the ones of the unperturbed states for $\eps>m \omega$. This labeling is arbitrary as long as one stays consistent throughout the calculation.}. They are defined below.  In the far off-resonant case,  $\ketfl{\Phi_{+,n+m}^\text{RWA}}$ corresponds for $\eps>m \omega$  to the unperturbed state $\uPlus{n+m}$, and  $\ketfl{\Phi_{-,n}^\text{RWA}}$ to $\uMin{n}$. For $\eps< m \omega$,  the state $\ketfl{\Phi_{+,n+m}^\text{RWA}}$ corresponds to $\uMin{n}$, and $\ketfl{\Phi_{-,n}^\text{RWA}}$ to $\uPlus{n+m}$.  The eigenenergies are 
\begin{align}
 \hbar \eps_{-,n}^\text{RWA} &= \hbar [ (-n - \half m) \omega - \half \Omega^\text{RWA}_m], \\
 \hbar \eps_{+,n+m}^\text{RWA} &= \hbar [ (-n - \half m) \omega + \half \Omega^\text{RWA}_m]
\end{align} 
with  the  oscillation frequency
\begin{equation} \label{FreqRWA}
   \Omega^\text{RWA}_m \equiv   \sqrt{(-\eps + m \omega)^2 + \DDelta{-m}^2}. 
\end{equation} 
The corresponding eigenstates are
\begin{align}
  \PflMin{n}{\text{RWA}} &= -\sinmix{\text{RWA}} \uMin{n} \nonumber \\
                       &- \sign{\Dm} \cosmix{\text{RWA}} \uPlus{n+m}, \label{FiMinQD} \\
  \PflPlus{n+m}{\text{RWA}} &= \cosmix{\text{RWA}} \uMin{n} \nonumber \\
              &- \sign{\Dm} \sinmix{\text{RWA}} \uPlus{n+m}  \label{FiPlusQD},
\end{align} 
where
\begin{equation} \label{Theta}
  \tan \Theta_{m}^\text{RWA} =  \frac{|\Dm|}{-\eps +m \omega} \quad \text{for} \quad 0 < \Theta_{m}^\text{RWA} \leq \pi.
\end{equation}
Now we are able to recover the time-dependent dynamics of the system (see  Appendix \ref{app::dynamics}). As an example, we give the state returning probability for a system starting in the localized state $\ket{\downarrow}$ and returning to this state:
\begin{align} \label{PSurQD}
  P_{\downarrow \rightarrow \downarrow}^\text{RWA} (t) &= \cos^2 \left(\Omega^\text{RWA}_m \frac{t}{2}\right) + \cos^2 \Mixang{\text{RWA}} \sin^2 \left(\Omega^\text{RWA}_m \frac{t}{2}\right).
\end{align} 
For the special case of vanishing static bias ($\eps = 0$ ) and 0-photon resonance,
\begin{equation} \label{ZeroPhotonRes}
 P_{\downarrow \rightarrow \downarrow}^\text{RWA} (t)= \cos^2 \left(|J_0 (A/\omega) \Delta| \frac{t}{2}\right),
\end{equation} 
which agrees with the high-frequency result, $\omega \gg \Delta$, of earlier works \cite{Shirley1965, Grifoni1998, Goychuk2005}.

\subsection{Van Vleck perturbation theory} \label{sec::VanVleck}
As pointed out already in \cite{Oliver2005, Son2009}, the RWA fails in explaining higher order effects in $\Delta$ such as a shift in the oscillation frequency. Furthermore, we will show that the couplings between the nondegenerate states in (\ref{FloquetMatrix}) are needed to get physically correct expressions for the relaxation and dephasing rates.  In the following, we will use Van Vleck perturbation theory to go beyond those shortcomings. Originally this method was used to treat modifications on  diatomic molecules caused by vibrations and rotations of the nuclei \cite{VanVleck1929}. Since then the formalism has found many applications in both chemistry and physics and experienced various modifications; see, for example,  \cite{Kirtman1968, Certain1970, Shavitt1980, Kirtman1981}. The main formalism behind these different variants is, however, always the same: a unitary transformation $U$ is applied in order  to construct an effective Hamiltonian which exhibits, to a certain order in the perturbation, the same eigenenergies as the original Hamiltonian but only connects almost degenerate levels.  In this work we choose for the transformation the form $U=\exp(\rmi S)$, which was originally proposed by Kemble in \cite{Kemble1937} and is described in more detail in \cite{CohenTannoudji2004}. In the case of the Floquet Hamiltonian, the effective Hamiltonian then becomes: $\fh_\text{eff} = \exp(\rmi S) \fh_\text{TLS} \exp(-\rmi S)$. We calculate the transformation matrix $S$ up to second-order in $\Delta$. As shown in Appendix \ref{app::VanVleck} the so-obtained effective Hamiltonian for an $m$-photon resonance again consists of 2 $\times$ 2 blocks, however, compared to the one of the previous section, it has corrected diagonal entries:
\begin{equation} \label{Heff2}
  \left( \begin{array}{c c}
  \hbar \epsMin{n} - \frac{\hbar}{4} \sum_{l \neq -m} \frac{|\DDelta{l}|^2}{\eps + l \omega} & -\hh \DDelta{-m} \\
  -\hh \DDelta{-m} & \hbar \epsPlus{n+m} + \frac{\hbar}{4} \sum_{l \neq -m} \frac{|\DDelta{l}|^2}{\eps + l \omega}
\end{array}
  \right).
\end{equation}
It leads to the new quasienergies 
\begin{align}
 \hbar \eps_{-,n}  &= \hbar [(-n - \half m) \omega - \half \Omega_m^{(2)}], \label{QuasiEnerVVMin} \\
 \hbar \eps_{+,n+m}  &= \hbar [(-n - \half m) \omega + \half \Omega_m^{(2)}], \label{QuasiEnerVVPlus} 
\end{align} 
 with the  second-order oscillation frequency \footnote{We wish to point out that we perform the calculation of the effective Hamiltonian (\ref{Heff2}) and the corresponding transformation matrix only to second-order in $\Delta$, whereas for the frequency [Eq. (\ref{FreqVV})], the mixing angle [Eq. (\ref{Theta2})],  and  the calculation of the survival probability, we retain also higher orders.}
\begin{equation} \label{FreqVV}
  \Omega_m^{(2)}  =  \sqrt{\left( -\eps + m \omega -\half \sum_{l \neq -m} \frac{|\DDelta{l}|^2}{\eps + l \omega} \right)^2 + \DDelta{-m}^2}.
\end{equation}
Compared to the frequency obtained within the RWA, Eq. (\ref{FreqRWA}), this new frequency is shifted due to the second-order elements in (\ref{Heff2}), and the condition for an $m$-photon resonance reads now
\begin{equation} \label{ResCondVV}
 \eps = m \omega -\half \sum_{l \neq -m} \frac{|\DDelta{l}|^2}{\eps + l \omega}.
\end{equation} 
\begin{figure}[h]
 \includegraphics[width=.45\textwidth]{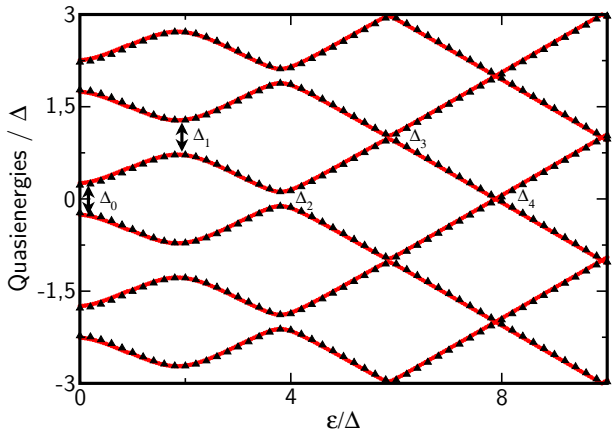}
 \caption{(Color online) Quasienergies against static bias $\eps$. The upward triangles result from numerical diagonalization of the Floquet matrix (\ref{FloquetMatrix}), while the solid  lines correspond to the analytical formulas (\ref{QuasiEnerVVMin}) and (\ref{QuasiEnerVVPlus}). Parameters are $\omega/\Delta =2$, $A/\Delta =3$. At an $m$-photon resonance, we find avoided crossings with a gap distance of $\DDelta{m}$. \label{fig::Quasienergies}}
\end{figure}
In Fig. \ref{fig::Quasienergies}, we compare Eqs. (\ref{QuasiEnerVVMin}) and (\ref{QuasiEnerVVPlus}) for the quasienergies against the eigenenergies we find from numerical diagonalization of the Floquet matrix (\ref{FloquetMatrix}). Whenever the resonance condition [Eq. (\ref{ResCondVV})] is fulfilled, we notice avoided crossings whose gap distance is determined by $\DDelta{m}$ for an $m$-photon resonance.
 The eigenstates $\ketfl{\Phi_{\pm,n}^\text{eff}}$  of the effective Hamiltonian  are the same as in (\ref{FiMinQD}) and (\ref{FiPlusQD}), with the mixing angle $\Mixang{\text{RWA}}$ replaced by
\begin{equation} \label{Theta2}
   \Theta_{m} = \arctan \left(\frac{|\DDelta{-m}|}{-\eps + m \omega -\frac{1}{2} \sum_{l \neq -m} \frac{|\DDelta{l}|^2}{\eps + l \omega} } \right).
\end{equation}
To get the eigenstates of $\fh_\text{TLS}$, we calculate $\ketfl{\Phi_{\pm,n}} = \exp(-\rmi S)\ketfl{\Phi_{\pm,n}^\text{eff}} $ and, following Appendix \ref{app::dynamics}, the survival probability $P_{\downarrow \rightarrow \downarrow}(t)$.
\begin{figure}
 \includegraphics[width=.45\textwidth]{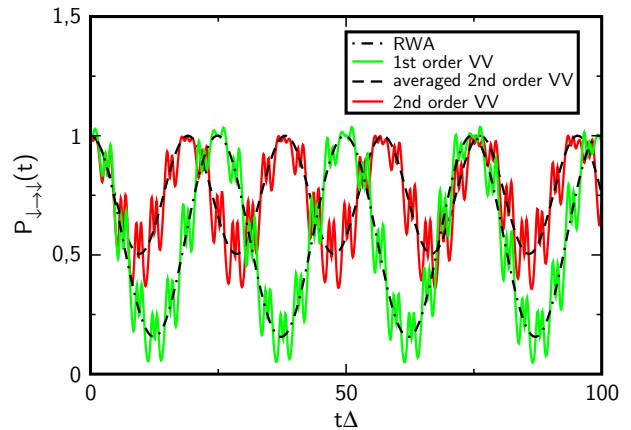}
 \caption{(Color online) Survival probability $P_{\downarrow \rightarrow \downarrow}(t)$ close to a 2-photon resonance. The parameters are $\eps / \Delta = 4.1$, $\omega / \Delta= 2 $, $A / \Delta = 3$. We compare results obtained from a RWA and the first- and second-order Van Vleck perturbation theory. Furthermore, we show the averaged second-order Van Vleck dynamics, which correspond to Eq. (\ref{PSurQD}) with the Van Vleck frequency [Eq. (\ref{FreqVV})] and second-order mixing angle [Eq. (\ref{Theta2})]. \label{fig::PNonDiss}}
\end{figure}
 In Fig. \ref{fig::PNonDiss}, we visualize the results for the survival probability  close to a 2-photon resonance obtained from the RWA approach and first- and second-order Van Vleck perturbation theory. We notice that by applying the RWA the fast oscillations in the first-order Van Vleck result are averaged out. When we compare first- and second-order predictions, the shift of the oscillation frequency is striking. But also the amplitude of the oscillations changes, which is due to the corrected mixing angle, Eq. (\ref{Theta2}). Inserting the second-order mixing angle and  frequency into the RWA formula (\ref{PSurQD}) results in averaging over the fast oscillations of the second-order Van Vleck graph. To also cover the fast driving-induced oscillations,  it is essential to use the eigenstates $\ketfl{\Phi_{\pm,n}}$ instead of the effective ones, which leads to a more complicated formula for $P_{\downarrow \rightarrow \downarrow}(t)$, see Appendix \ref{app::dynamics}.

\subsection{Validity of the Van Vleck approach} \label{sec::ValidVanVleck}
In closing this section, we give an overview of the parameter regime in which our approach is valid.
To apply Van Vleck perturbation theory at all, a requirement for the Floquet Hamiltonian is that it has  for finite $\Delta$ a similar doublet structure as in the unperturbed case ($\Delta=0$). This means that the off-diagonal elements in (\ref{FloquetMatrix}) connecting different doublets with each other must be much smaller than the distance between those doublets \cite{CohenTannoudji2004}:
\begin{equation}
   |\brafl{u_{\uparrow,n}^0} \Delta \sigma_x \ketfl{u_{\downarrow,l+m+n}^0}| \ll |\eps_{\uparrow,n}^0 - \eps_{\downarrow,l+m+n}^0|
\end{equation} 
for any $l\neq0$.
Using  Eqs. (\ref{DressedDelta}) and (\ref{ResCond}), this becomes
\begin{equation} \label{VanVleckCondGen}
 |\Delta_{-l-m}| \ll |\eps - (l+m) \omega|.
\end{equation} 
Because $|\Delta_{-l-m}| \leq \Delta$, this condition can be even fulfilled for frequencies $\omega < \Delta$. Once Eq. (\ref{VanVleckCondGen}) is valid, we still have to check at which order one can stop the perturbative expansion in $\Delta$.
We will distinguish now between two situations: the case of being close to or at an $m$-photon resonance and the regime far from resonance.

\subsubsection{Dynamics close to or at resonance}
\begin{figure}[h]
 \includegraphics[width=.48\textwidth]{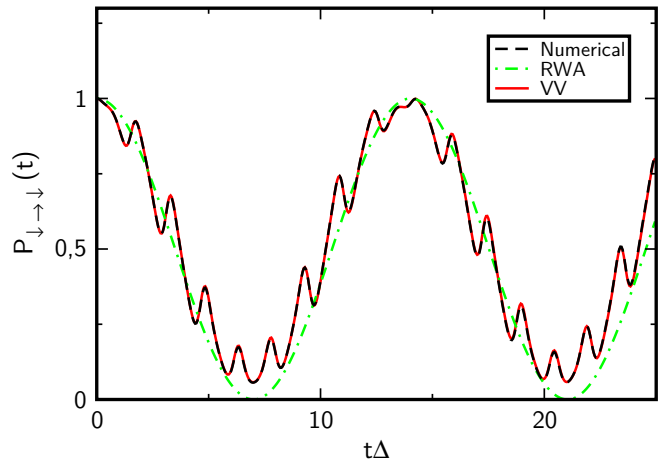}
 \caption{(Color online) Survival probability $P_{\downarrow \rightarrow \downarrow}(t)$ for $\eps/\Delta = 4$, $\omega/\Delta = 4$, and $A/\Delta =4.1$. Exact numerical results are compared with RWA and second-order Van Vleck results. \label{fig::PAtResonance}}
\end{figure}

\begin{figure}[h]
 \includegraphics[width=.48\textwidth]{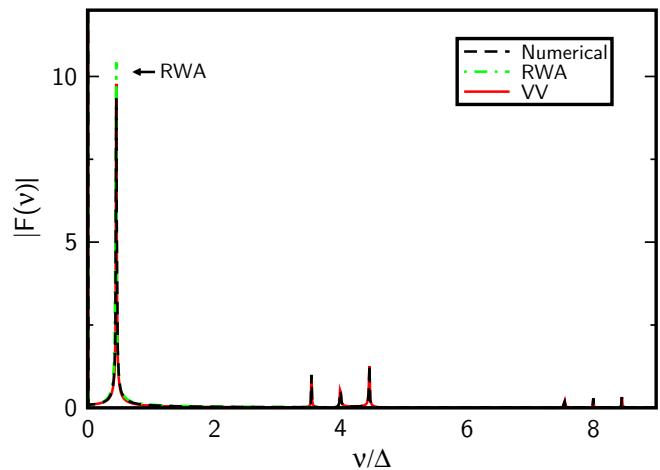}
 \caption{(Color online) Absolute value of the Fourier transform $F(\nu)$ of $P_{\downarrow \rightarrow \downarrow}(t)$ in Fig. \ref{fig::PAtResonance}. The oscillation frequency corresponding to  $\Omega_1  = 0.45 \Delta$ is dominating and is also predicted by the RWA approach. Notice, that there is exact agreement between the second-order Van Vleck and the numerical results. \label{fig::FAtResonance}}
\end{figure}

Using $\eps \sim m \omega$, Eq. (\ref{VanVleckCondGen}) becomes simply
\begin{equation} \label{VanVleckCond}
  \omega \gg  \frac{|\DDelta{-l-m}|}{|l|}.
\end{equation}
Notice that the right-hand side of (\ref{VanVleckCond}) still depends on $A/\omega$. Thus,  while being surely fulfilled in the RWA case, $\omega \gg \Delta$, condition (\ref{VanVleckCond}) is  in general less restrictive. To show this, we examine the following two limiting cases. First, the limit $A / \omega \ll 1$ is considered. For arguments with $0 < x \ll \sqrt{n+1}$, the $n$th-order Bessel function becomes approximately \cite{Arfken2001}
\begin{equation}
  |J_{n}(x)| \approx \frac{x^{|n|}}{2^{|n|} |n|!}.
\end{equation} 
Thus, for $A/\omega \ll 1$, we find that
\begin{equation}
  |\Delta_n| \approx \frac{(A/\omega)^{|n|}}{2^{|n|} |n|!} \Delta
\end{equation}  
and (\ref{VanVleckCond}) becomes
\begin{equation} \label{VanVleckCondSmallArg}
  \omega \gg \Delta \frac{(A/\omega)^{|-l-m|} }{|l|}.
\end{equation} 
Because $A/\omega \ll 1$, Eq. (\ref{VanVleckCondSmallArg}) is fulfilled for any $l\neq 0$ if it is satisfied for $l=-m$; i.e., if
\begin{equation} \label{VanVleckCondSmallArg2}
  \omega \gg \frac{\Delta}{|m|}.
\end{equation} 
In the case of a 1-photon resonance, this leaves us with the RWA condition, $\omega \gg \Delta$, as then nearest neighbor doublets are connected by a $\DDelta{0}$ element in the Floquet matrix which approaches $\Delta$ for small $A/\omega$. All other perturbative off-diagonal entries in (\ref{FloquetMatrix}) are vanishingly small. In the case of an $m$-photon resonance with $m \neq \pm 1$, the dressed element $\DDelta{0}$ connects more distant doublets, so that the Van Vleck condition (\ref{VanVleckCond}) can be realized according to (\ref{VanVleckCondSmallArg2}) for  frequencies smaller than the ones demanded by the RWA.\newline
In the opposite limit of $A/\omega \gg 1$, an upper bound for the dressed Bessel function is \cite{Arfken2001}
\begin{equation}
  |\DDelta{n}| \leq \Delta \sqrt{\omega/A}.
\end{equation} 
Using this,  we find that (\ref{VanVleckCond}) is verified if
\begin{equation} \label{VanVleckAsymp}
  \omega \gg \frac{\Delta^2}{A}.
\end{equation} 
Since $A \gg \omega$, it follows that $\Delta^2 \ll A \omega < A^2$ and thus $A \gg \Delta$. Hence, Eq. (\ref{VanVleckAsymp}) represents an improvement to the RWA condition.\newline
Further, being close to an $m$-photon resonance, one single frequency will dominate the system's behavior, and thus neglecting the remaining fast-oscillating terms will already give a good picture of the coarse-grained dynamics. This dominating frequency is  represented by $\Omega_m^\text{RWA}$ and by $\Omega_m^{(2)}$ in the case of the RWA and the second-order Van Vleck perturbation theory, respectively. To obtain those frequencies, it is enough to diagonalize the corresponding effective Hamiltonian, without yet considering any modification of the eigenstates of the effective Hamiltonian. As shown in the previous subsection, $\Omega_m^\text{RWA}$ corresponds to the main frequency of the system obtained by applying Van Vleck perturbation theory to first order in $\Delta$. Naturally the question arises as to how good these approximations are, or which orders in $\Delta$ are necessary depending on the parameter regime.\newline
In a first step, we examine the improvement obtained by using second-order Van Vleck perturbation theory compared to the RWA; that is, we consider
\begin{figure}[h]
 \includegraphics[width=.48\textwidth]{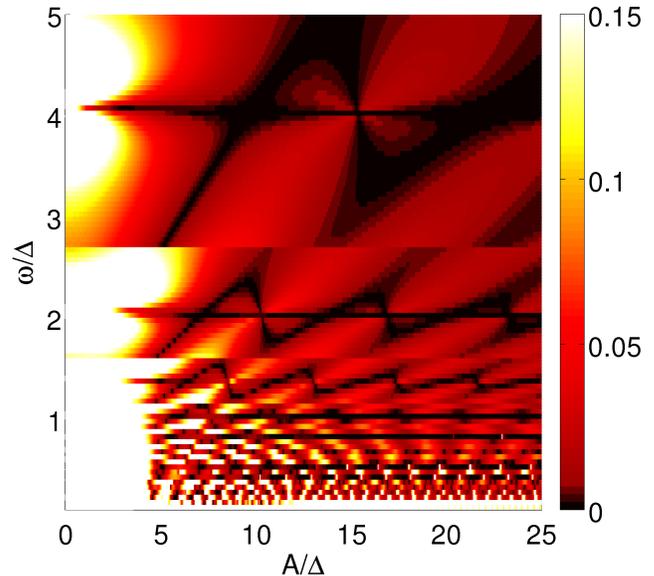}
 \caption{(Color online) Comparison of the main oscillation frequency $\Omega_m^\text{RWA}$ obtained by the RWA and second-order Van Vleck frequency $\Omega_m^{(2)}$ for a fixed static bias, $\eps = 4.0 \Delta$. The relative mistake performing the RWA is shown against the driving frequency $\omega /\Delta$ and driving amplitude $A/\Delta$. The darkest areas show regions in parameter space of small or no deviations between the two approaches, whereas the lightest areas show a deviation of 15 \% or more. \label{fig::SecondOrdCorr}}
\end{figure}
\begin{equation}
  \epsilon^\text{RWA} = \frac{|\Omega^\text{RWA}_m - \Omega_m^{(2)}|}{\Omega_m^{(2)}},
\end{equation} 
 and plot it in Fig. \ref{fig::SecondOrdCorr} against the driving frequency $\omega$ and amplitude $A$ at a fixed value of the static bias $\eps$. The deviations are visualized through different shades of color.  The lightest areas stand for a relative mistake of 15 \% or more. We can tell from Fig. \ref{fig::SecondOrdCorr} that the RWA fails for low driving frequencies and/or weak driving amplitudes.
The darkest areas determine regions in the parameter space where almost no difference between the RWA and second-order Van Vleck approach can be found. Of course this is no indication that the results are reliable in those areas, but rather that second-order perturbation theory yields no improvement to the RWA.\newline
 \begin{figure}[h]
 \includegraphics[width=.48\textwidth]{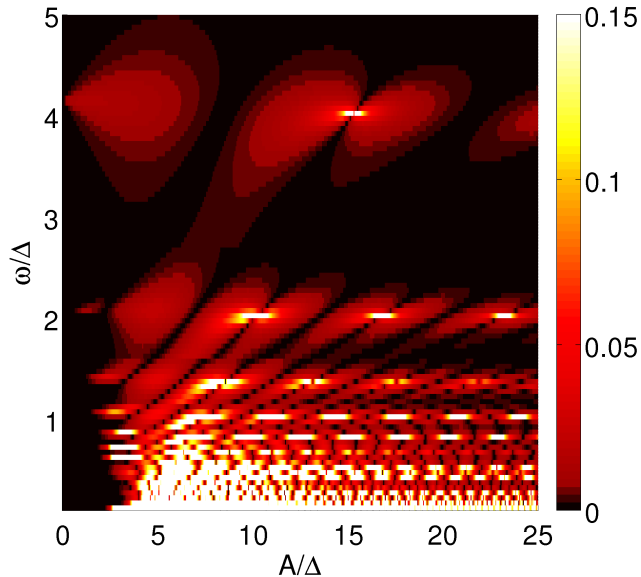}
 \caption{(Color online) Comparison of the main oscillation frequency $\Omega_m$ obtained by  second-order and third-order Van Vleck perturbation theory for a fixed static bias, $\eps = 4.0 \Delta$. The relative mistake performing second-order perturbation theory is shown against the driving frequency $\omega /\Delta$ and driving amplitude $A/\Delta$. Color scale is the same as in Fig. \ref{fig::SecondOrdCorr}. \label{fig::ThirdOrdCorr}}
\end{figure}
To check the accuracy of the second-order Van Vleck frequency $\Omega_m^{(2)}$, we calculate the deviation
\begin{equation}
  \epsilon^\text{(2)} = \frac{|\Omega_m^{(2)}-\Omega_m^{(3)}|}{\Omega_m^{(3)}}
\end{equation} 
from the frequency $\Omega_m^{(3)}$ obtained applying Van Vleck perturbation theory to third order \cite{Son2009}. Results for $\epsilon^\text{(2)}$ are shown in Fig. \ref{fig::ThirdOrdCorr}.
Again we only consider mistakes up to 15 \%. We find strong deviations in the region of low driving frequency and intermediate driving amplitudes. In the remaining parameter space, the agreement between second- and third-order Van Vleck perturbation theory is quite good apart from small islands. Those islands are located at values of $\omega$ and $A$ where the second-order condition for coherent destruction of tunneling (CDT) is fulfilled, see discussion in Sec. \ref{sec::CDT}. For example, for $\eps / \Delta =4.0 $ and $\omega /\Delta =2.0$, they occur at the zeros of the Bessel function $J_{2}(A/\omega)$. Since at those points the second-order frequency $\Omega_m^{(2)}$ vanishes, even small third-order contributions yield a significant correction. This behavior visualizes nicely the findings of Barata \textit{et al.}  \cite{Barata2000} and Frasca \cite{Frasca2005}, who proved analytically that $\Omega_m$ does not completely vanish at the zeros of the Bessel function if third-order contributions in $\Delta$ are taken into account. On the contrary, both the RWA and second-order Van Vleck perturbation theory predict a vanishing frequency at those points and therefore agree perfectly with each other in Fig. \ref{fig::SecondOrdCorr}.  We want to emphasize again that, as can be seen from Fig. \ref{fig::ThirdOrdCorr}, our approach also yields  good results for low driving frequencies, $\omega < \Delta$, and small driving amplitudes, $A \sim \Delta$.\newline
In Figs. \ref{fig::PAtResonance} and \ref{fig::FAtResonance}, we show the survival probability $P_{\downarrow \rightarrow \downarrow}(t)$ and its Fourier transform
\begin{equation}
  F(\nu) := \int_{-\infty}^\infty dt \, P_{\downarrow \rightarrow \downarrow}(t) \rme^{\rmi \nu t} 
\end{equation} 
 at resonance $\eps = \omega$ but for a driving amplitude with $|J_1(A/\omega)| \neq 0$. One clearly sees that one frequency, namely, $\Omega_1$, is dominating, and already the RWA conveys a good impression of the dynamics.

\subsubsection{Dynamics away from resonance}
\begin{figure}[h]
 \includegraphics[width=.48\textwidth]{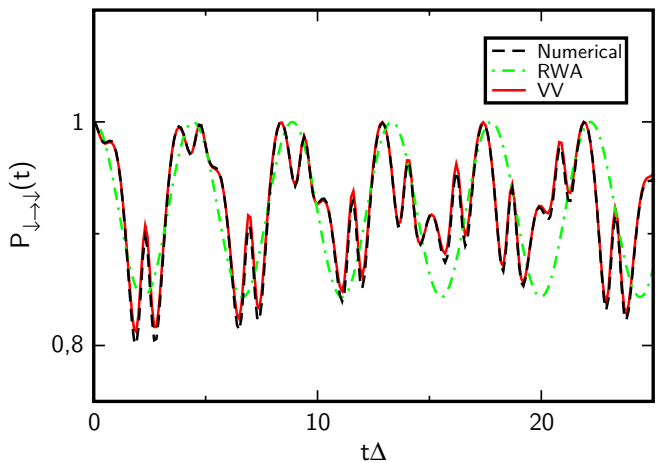}
 \caption{(Color online) Survival probability $P_{\downarrow \rightarrow \downarrow}$  for $\eps/\Delta = 4$, $\omega/\Delta = 2.7$, and $A/\Delta =4.1$. Exact numerical results are compared with RWA and second-order Van Vleck predictions. \label{fig::PBetweenResonances}}
\end{figure}

\begin{figure}[h]
 \includegraphics[width=.48\textwidth]{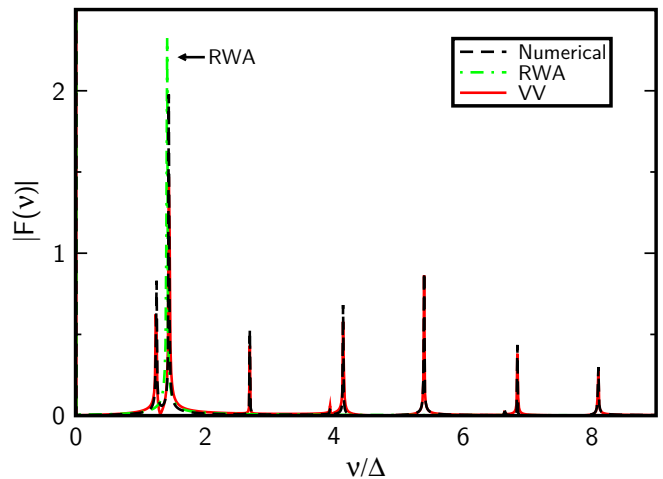}
 \caption{(Color online) Absolute value of the Fourier transform of $P_{\downarrow \rightarrow \downarrow}(t)$ in Fig. \ref{fig::PBetweenResonances}. The numerical and the second-order Van Vleck graphs clearly show several dominating oscillation frequencies, while the RWA only shows one of them. Note also the different scale of the $y$ axis compared to Fig. \ref{fig::FAtResonance}. \label{fig::FBetweenResonances}}
\end{figure}

The situation changes when we are away from a resonance. Already intuitively it becomes clear that the dynamics will not be governed anymore by a single frequency. Therefore, by looking only at the coarse-grained dynamics of the system and averaging out the driving-induced oscillations,  significant information  is lost. This case is presented in Figs. \ref{fig::PBetweenResonances} and \ref{fig::FBetweenResonances}, where we are in the region between the 1- and 2-photon resonances. In contrast to Figs. \ref{fig::PAtResonance} and  \ref{fig::FAtResonance}, we find that several frequencies are dominating and determine the dynamics of the system. The second-order Van Vleck approach reflects this behavior almost perfectly because the driving-induced oscillations are accounted for. However, the RWA shows only  one single  oscillation because the others  are averaged out. It depends on the choice of $m$ in the formula for the RWA, Eq. (\ref{FreqRWA}), which of the frequencies is taken. This explains also the cuts in Fig. \ref{fig::SecondOrdCorr}; see, for example, the horizontal line just below $\omega/\Delta \approx 3$. At these values of the frequency, we change $m$ in our analytical calculation. In Fig. \ref{fig::ThirdOrdCorr}, those cuts are barely visible. Being away from the resonance point, the modifications of the external driving on the system's eigenstates must not be neglected.\newline
Off resonance, the requirement (\ref{VanVleckCondGen}) for Van Vleck perturbation theory  is surely fulfilled for a large enough static bias,
\begin{equation}
 |\DDelta{-l-m}| \ll \eps. 
\end{equation}

\section{The dissipative system}\label{sec::DissSys}
To include dissipative effects on our system, we consider the time-dependent spin-boson Hamiltonian \cite{Weiss2008, Grifoni1998,Goychuk2005}
\begin{equation} \label{DissHam}
   H(t) = H_\text{TLS} (t) + H_\text{B} + H_\text{int},
\end{equation} 
where the environmental degrees of freedom are mo\-deled by an infinite set of harmonic oscillators, $H_\text{B} = \sum_k \hbar \nu_k b^\dagger_k b_k$, which are bilinearly coupled to the TLS by the coupling Hamiltonian
\begin{equation}
 H_\text{int} = x \sum_k \hbar \lambda_k (b_k^\dagger+b_k) + x^2 \sum_k \hbar \frac{\lambda_k^2}{\nu_k}.
\end{equation} 
Here $x=\sigma_z/2$ is the position matrix of the TLS and $\lambda_k$ the coupling strength to the $k$th mode of the bath. The spectral density of the bath can be expressed as  $G(\nu) = \sum_k \lambda_k^2 (\nu -\nu_k)$. We assume further that at time $t=0$ the bath is in thermal equilibrium and uncorrelated to the system, so that  the full density matrix $W(t)$ associated with $H(t)$ has at initial time the form $W(0) = \rho(0) \otimes \rho_\text{B}(0)$, where $\rho(t)$ is the density matrix of the TLS and $\rho_\text{B}(0) = \exp(-\beta H_\text{B}) / \text{tr}_\text{B} \exp(-\beta H_\text{B})$ is the density matrix of the bath at temperature $T = (k_B \beta)^{-1}$.  Following \cite{Blum1996, Louisell1973, Kohler1997} and performing a Born and Markov approximation, we arrive at the Floquet-Bloch-Redfield master equation
\begin{equation} \label{FloqMaster}
  \dot \rho_{\alpha \beta} (t) = - \rmi (\eps_\alpha - \eps_\beta) \rho_{\alpha \beta} (t) + \pi \sum_{\alpha^\prime, \beta^\prime} \LDiss{\alpha \beta}{\alpha^\prime \beta^\prime}(t) \rho_{\alpha^\prime \beta^\prime}(t), 
\end{equation} 
where the density matrix is expressed in the basis of the energy eigenstates of the TLS:
\begin{equation}
  \rho_{\alpha \beta}(t) = \bra{\Phi_\alpha(t)} \rho(t) \ket{\Phi_\beta (t)},  \qquad \alpha, \beta = \pm. 
\end{equation}
Notice that $\eps_\alpha \equiv \eps_{\alpha, 0}$ and $\ket{\Phi_\alpha (t)} \equiv \ket{\Phi_{\alpha,0}(t)}$. Corrections to the oscillation frequencies due to the Lamb shift are not accounted for. The first part of (\ref{FloqMaster}) describes the nondissipative dynamics as treated in Sec. \ref{sec::NonDiss}. The influence of the bath is fully characterized by the time-dependent rate coefficients
 \begin{align} \label{Coeff}
  &\LDiss{\alpha\beta}{\alpha^\prime \beta^\prime}(t) = \nonumber \\
     & \sum_{n, n^\prime} \rme^{\rmi (n+n^\prime) \omega t} \biggl\{ (N_{\alpha \alpha^\prime, n} + N_{\beta \beta^\prime, -n^\prime}) \X{\alpha \alpha^\prime}{n} \X{\beta^\prime \beta}{n^\prime} \nonumber \\
                           &- \delta_{\beta \beta^\prime} \sum_{\beta^{\prime \prime}} \X{\alpha \beta^{\prime \prime}}{n^\prime} N_{\beta^{\prime \prime} \alpha^\prime,n} \X{\beta^{\prime \prime}\alpha^\prime}{n} \nonumber \\
                           &- \delta_{\alpha \alpha^\prime} \sum_{\alpha^{\prime \prime}}  N_{\alpha^{\prime \prime} \beta^\prime,-n} \X{\beta^\prime \alpha^{\prime \prime}}{n} \X{\alpha^{\prime \prime}\beta}{n^\prime} \biggr\}
\end{align}
with $N_{\alpha \beta,n} = N(\eps_\alpha - \eps_\beta +  n \omega)$, $N(\nu) = G(\nu) n_\text{th}(\nu)$, and $n_\text{th} (\nu) = \frac{1}{2} \left[ \coth (\hbar \beta \nu/2) -1   \right]$. As also the matrix elements of the position operator $x =  \sigma_z /2$ are periodic in time, we express them in a Fourier series, $\bra{\Phi_\alpha(t)} x \ket{\Phi_\beta(t)} = \sum_n \rme^{\rmi n \omega t} \X{\alpha \beta}{n}$.

\subsection{Position matrix elements} \label{sec::PosMat} 
The Fourier coefficients appearing  in the rate equations (\ref{Coeff}) can be calculated by
\begin{align} \label{X}
  \X{\alpha \beta}{l} &= \frac{1}{T} \int_{0}^T dt \, \rme^{- \rmi l \omega t} \bra{\Phi_\alpha (t)} x \ket{\Phi_\beta (t)} \nonumber \\
                      &= \brafl{\Phi_{\alpha,0}} x \ketfl{\Phi_{\beta,l}},
\end{align}
where we used the periodicity of the eigenfunctions of the TLS and the definition of the internal product in the composite Hilbert space, Eq. (\ref{InternalProduct}). From this we find that $\X{\alpha \beta}{-n} = \X{\beta \alpha}{n}^* $ and that we can use the Floquet eigenstates  (\ref{app::FiMin2}) and (\ref{app:FiPlus2}) to calculate the Fourier coefficients to second-order in $\Delta$. We get
\begin{align}
  \X{-+}{n}^{(2)} &= \X{-+}{n}^{(1)}(\Mixangle) \nonumber \\
       &+ \frac{\sin \Mixangle}{8}  \sum_{k\neq n, m} \frac{\DDelta{n-k-m}\DDelta{-k}}{[\eps+(n-k-m)\omega][-\eps+k\omega]}, \label{X-+2}
\end{align} 
\begin{align}
  \X{--}{n}^{(2)} &= \X{--}{n}^{(1)} (\Mixangle) \nonumber \\
             &- \frac{\cos \Mixangle}{8}  \sum_{k\neq n+m, m} \frac{\DDelta{n-k}\DDelta{-k}}{[\eps+(n-k)\omega][-\eps+k\omega]}, \label{X--2}
\end{align}
with 
\begin{align}
  \X{-+}{n}^{(1)} (\xi) &=   \frac{\sin \xi}{2}  \delta_{n,m} -  \frac{\sign{\Delta_{-m}}}{2}  \biggl[ \sin^2 \frac{\xi}{2} \frac{\DDelta{-n}}{-\eps + n \omega} \nonumber \\
            &  + \cos^2 \frac{\xi}{2} \frac{\DDelta{n-2m}}{\eps +(n-2m) \omega}\biggr] (1- \delta_{n,m}) \label{X-+1},
\end{align}
\begin{align}
  \X{--}{n}^{(1)} (\xi) &= - \frac{\cos \xi}{2}  \delta_{n,0}+  \frac{\sign{\DDelta{-m}}}{4}  \sin \xi   \nonumber\\
  &  \biggl[ \frac{\DDelta{-m-n}}{-\eps + (m + n) \omega} -  \frac{\DDelta{n-m}}{\eps +(n-m) \omega}\biggr] (1- \delta_{n,0}), \label{X--1}
\end{align}
where for $\xi$ either the mixing angle $\Theta_m^\text{RWA}$ or $\Theta_m$ is used.
Further, we find that $\X{++}{n}^{(2)} = - \X{--}{n}^{(2)}$. Within the RWA, we would get $\X{-+}{n}^\text{RWA} = \half \sin \Mixangle^\text{RWA} \delta_{n,m}$ and $\X{--}{n}^\text{RWA} = -\half \cos \Mixangle^\text{RWA} \delta_{n,0}$. From this we notice that, in the case of a simple RWA, $\X{-+}{n}^\text{RWA}$ would be nonzero  for $n=m$ and $\X{--}{n}^\text{RWA}$ for $n=0$ only. An improvement to that can already be achieved by using Van Vleck perturbation theory to first order in $\Delta$, yielding $\X{\alpha \beta}{n}^{(1)} (\Mixangle^\text{RWA})$. It contains next to the RWA results additionally first-order corrections for any index $n$ in $\X{\alpha \beta}{n}$.
\begin{figure}[h]
 \includegraphics[width=.45\textwidth]{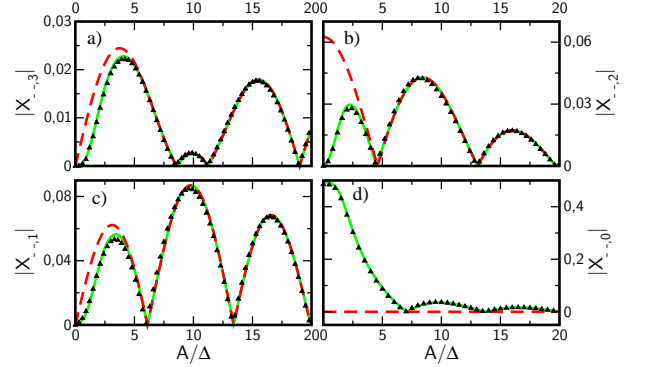}
 \caption{(Color online) Fourier coefficient $|\X{--}{n}|$ for various values of $n$ against driving amplitude $A$. We examine the case $\eps=2 \omega$. The parameters are $\eps /\Delta= 4.0 $ and $\omega / \Delta = 2.0$. The black triangles show data points from numerical diagonalization of the Floquet matrix; the red (dark gray) dashed curve is obtained from first-order perturbation in $\Delta$ [Eq. (\ref{X--1})], whereas the green (light gray) solid curve is obtained by going to second-order in $\Delta$ [Eq. (\ref{X--2})]. \label{fig::X--}}
\end{figure}
\begin{figure}[h]
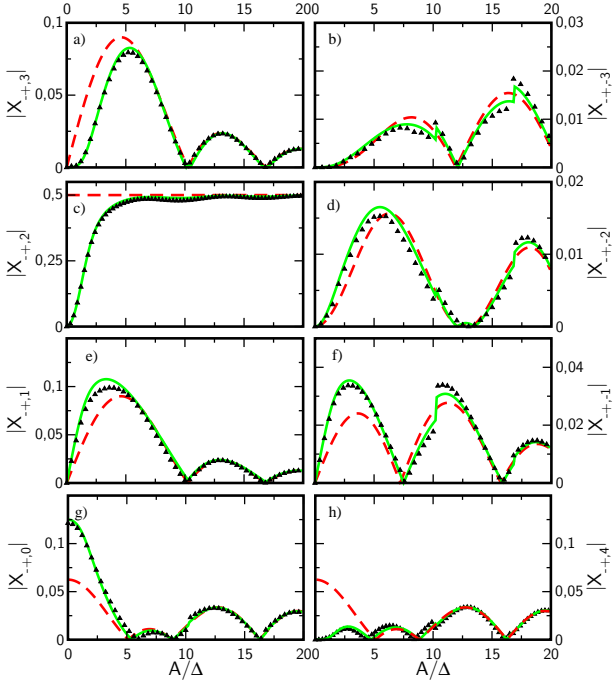

 \includegraphics[width=.45\textwidth]{Figures/Fig11_a-d.eps}
 \includegraphics[width=.45\textwidth]{Figures/Fig11_e-f.eps}
 \includegraphics[width=.45\textwidth]{Figures/Fig11_g-h.eps}
 \caption{(Color online) Fourier coefficient $|\X{-+}{n}|$ for various values of $n$ against driving amplitude $A$. The parameters are the same as in Fig. \ref{fig::X--}. \label{fig::X-+}}
\end{figure}
Figures \ref{fig::X--} and \ref{fig::X-+} show the absolute value of the coefficients $\X{--}{n}$ and $\X{-+}{n}$, respectively. We find a good agreement between the results obtained by a numerical diagonalization of the Floquet matrix (\ref{FloquetMatrix}) and second-order Van Vleck perturbation theory, Eqs. (\ref{X-+2}) and (\ref{X--2}). Concerning Figs. \ref{fig::X--}(b) and (d) we see  a qualitative improvement by going from first to second-order in $\Delta$. While in Fig. \ref{fig::X--}(b) the first-order result approaches a nonvanishing coefficient $\X{--}{2}$ for $A \rightarrow 0$, Eq. (\ref{X--2}) corresponds to the numerical calculation very well even in the region of low driving amplitude and meets our expectation that all Fourier coefficients except for $\X{\alpha \beta}{0}$ vanish at zero driving. The problem of the first-order results at low driving strength is caused by the definition of the first-order mixing angle $\Mixangle^\text{RWA}$, Eq. (\ref{Theta}), which is  $\pi/2$ for $\eps=m \omega$.  When $n \neq \pm m$ in $\X{--}{n}^{(1)} (\Mixangle^\text{RWA})$, the coefficient approaches zero for $A \rightarrow 0$ because of the term  
\begin{equation}
 \biggl[ \frac{\DDelta{-m-n}}{-\eps + (m + n) \omega} -  \frac{\DDelta{n-m}}{\eps +(n-m) \omega}\biggr]
\end{equation} 
in Eq. (\ref{X--1}). However, for $n=\pm m$ a zeroth-order Bessel function occurs in that part which does not vanish for $A \rightarrow 0$. A second-order improvement of the mixing angle as done in Eq. (\ref{Theta2}) solves this problem. In Fig. \ref{fig::X--}(d) the first-order solution predicts a coefficient $\X{--}{0}^{(1)} (\Mixangle^\text{RWA})$ which is constantly zero.\newline
Also, in Fig. \ref{fig::X-+} a noticeable improvement between first- and second-order perturbation theory can be seen. In \ref{fig::X-+}(c) the first-order solution shows a constant coefficient $\X{-+}{2}^{(1)} (\Mixangle^\text{RWA}) =0.5$.  We see from the numerics and second-order results that indeed the constant value is reached asymptotically for high driving amplitudes; however, for small driving amplitudes, we find a vanishing coefficient. 
 In Fig. \ref{fig::X-+}(h) we can observe a  behavior like in  \ref{fig::X--}(b), namely, that $\X{-+}{4}^{(1)} (\Mixangle^\text{RWA})$ does not approach zero for $A \rightarrow 0$. The explanation is similar to the above case.

\subsection{Moderate rotating-wave approximation} \label{sec::MRWA}
Having calculated the position matrix elements, our rate coefficients $\LDiss{\alpha \beta}{\alpha^\prime \beta^\prime}(t)$ are fully determined. What remains to do is to solve the Floquet-Bloch-Redfield master equation (\ref{FloqMaster}) for the density matrix $\rho$. For an analytical calculation, there is, however, still a difficulty: the time dependence of the coefficients. To get rid of this, we perform a moderate rotating-wave approximation (MRWA) \cite{Kohler1997}; i.e., we neglect fast-oscillatory terms in (\ref{Coeff}), which amounts to selecting only the terms with $n^\prime = -n$, and obtain
 \begin{align} \label{CoeffMRWA}
  &\LMRWA{\alpha\beta}{\alpha^\prime \beta^\prime} = \sum_{n}  \biggl\{ (N_{\alpha \alpha^\prime, n} + N_{\beta \beta^\prime, n}) \X{\alpha \alpha^\prime}{n} \X{\beta^\prime \beta}{-n} \nonumber \\
                           &- \delta_{\beta \beta^\prime} \sum_{\beta^{\prime \prime}} \X{\alpha \beta^{\prime \prime}}{-n} N_{\beta^{\prime \prime} \alpha^\prime,n} \X{\beta^{\prime \prime}\alpha^\prime}{n} \nonumber \\
                           &- \delta_{\alpha \alpha^\prime} \sum_{\alpha^{\prime \prime}}  N_{\alpha^{\prime \prime} \beta^\prime,-n} \X{\beta^\prime \alpha^{\prime \prime}}{n} \X{\alpha^{\prime \prime}\beta}{-n} \biggr\}.
\end{align}
We observe that $\LMRWA{\alpha \alpha}{-+} = \LMRWA{\alpha \alpha}{+-}$, $\LMRWA{-+}{\alpha \alpha} = \LMRWA{+-}{\alpha \alpha}$, $\LMRWA{\alpha \beta}{\alpha \beta} = \LMRWA{\beta \alpha}{\beta \alpha}$, and $\LMRWA{\alpha \beta}{ \beta \alpha} = \LMRWA{\beta \alpha}{ \alpha \beta}$. Moreover, $\rho_{--}(t)+\rho_{++}(t)=1$ and $\rho_{+-}(t) =\rho_{-+}^*(t)$. This yields simple expressions for the reduced density matrix elements to first order in the coupling $\kappa$ to the bath:
\begin{align}
  \rho_{--}(t) &= \pi \frac{\LMRWA{--}{++}}{\gamma_\text{rel}} + c_\text{rel} \frac{\rmi}{\pi} (m \omega + \Omega_m^{(2)}) \rme^{-\gamma_\text{rel}t} \nonumber\\
                &+ 2 \LMRWA{--}{-+} \text{Re}\{ c_\text{deph} \rme^{-\rmi (m \omega +  \Omega_m^{(2)})t} \} \rme^{-\gamma_\text{deph} t} \label{rho--},
\end{align}
\begin{align}
  \rho_{-+}(t) &= c_\text{rel} (\LMRWA{-+}{++}-\LMRWA{-+}{--})\rme^{-\gamma_\text{rel}t} \nonumber\\
               & + \half \LMRWA{-+}{+-} c_\text{deph} \rme^{-\rmi (m \omega +  \Omega_m^{(2)})t} \rme^{-\gamma_\text{deph} t} \nonumber\\
                & + c_\text{deph}^* \frac{\rmi}{\pi} (m \omega +  \Omega_m^{(2)}) \rme^{\rmi (m \omega +  \Omega_m^{(2)})t} \rme^{-\gamma_\text{deph} t} \label{rho-+}.
\end{align}
The constants $c_\text{rel}$ and $c_\text{deph}$ are fully determined by the initial conditions, Eq. (\ref{StartCond}).
The expressions for the relaxation and dephasing rates are
\begin{equation}
  \gamma_\text{rel} = \pi \left( \LMRWA{--}{++}- \LMRWA{--}{--} \right), \quad \gamma_\text{deph} = - \pi \LMRWA{-+}{-+}.
\end{equation}
 \begin{figure}[h]
 \includegraphics[width=.48\textwidth]{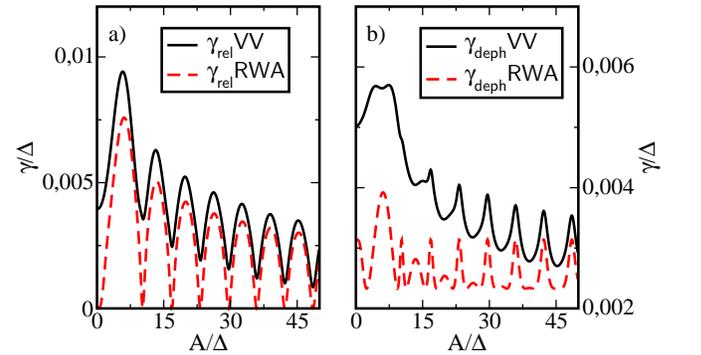}
 \caption{(Color online) Relaxation (a) and dephasing rate (b) against driving amplitude $A$ for $\omega /\Delta = 2.0$, $\eps / \Delta = 4.1$, $\hbar \beta \Delta =10$, and $\kappa = 0.01$. Results obtained within the second-order Van Vleck perturbation theory  are compared with RWA calculations. Notice that the RWA predicts an unphysical vanishing of the relaxation rates at the zeros of $\DDelta{-2}$. \label{fig::Rates}}
\end{figure}
With (\ref{CoeffMRWA}) we can express them in terms of the position matrix elements, yielding
\begin{equation}
 \gamma_\text{rel} = 4 \pi \sum_n \left[N_{-+,n} +\half \kappa (\eps_- - \eps_+ +  n \omega)   \right] \X{-+}{n}^2,
\end{equation} 
\begin{equation}
 \gamma_\text{deph} = \half \gamma_\text{rel} + 4 \pi \sum_n N_{--,n} \X{--}{n}^2.
\end{equation} 
With (\ref{X-+2}) and (\ref{X--2}) we arrive finally  at one major result:
\begin{equation}
  \gamma_\text{rel} = \gamma_\text{rel}^0 + \sum_{n \neq 0} \gamma_\text{rel}^n \quad \text{and} \quad \gamma_\text{deph} = \gamma_\text{deph}^0 + \sum_{n \neq 0} \gamma_\text{deph}^n 
\end{equation} 
with the contributions
\begin{align}
  \gamma_\text{rel}^0 &=  \pi G\left( \frac{\Omega_m^{(2)}}{2}\right) \coth \left(\frac{\hbar \beta}{2} \Omega_m^{(2)} \right) \sin^2 \Mixangle \nonumber \\
                       &\phantom{\mathrel{=}} \times \biggl[1 - \half \sum_{k \neq m} \frac{\DDelta{-k}^2}{(\eps - k\omega)^2} \biggr],
\end{align}
\begin{align}
     &\gamma_\text{rel}^n =  \pi  G \left[ \half (\Omega_m^{(2)}- n \omega) \right] \coth \left[\frac{\hbar \beta}{2} \left(  \Omega_m^{(2)}-  n \omega \right) \right] \nonumber \\
                  &  \times \left[ -\sin^2 \frac{\Mixangle}{2}\frac{\DDelta{-(n+m)}}{\eps - (n + m)\omega} + \cos^2 \frac{\Mixangle}{2}\frac{\DDelta{n-m}}{\eps + (n - m)\omega}   \right]^2
\end{align}
and
\begin{align}
  \gamma_{\rm deph}^0 &= \frac{1}{2} \gamma_{\rm rel} + \pi  N(0) \cos^2 \Mixangle  \biggl[1 - \half \sum_{k \neq m} \frac{\DDelta{-k}^2}{(\eps - k\omega)^2} \biggr], 
\end{align}
\begin{align}
 \gamma_\text{deph}^n &=   \frac{\pi}{8}  G(n \omega) \left[\coth \left( \frac{\hbar \beta}{2} n\omega \right) -1\right] \sin^2 \Mixangle \nonumber \\
      & \phantom{\mathrel{=}} \times \left[ \frac{\DDelta{-m-n}}{-\eps+(m+n)\omega} - \frac{\DDelta{n-m}}{-\eps+(n-m)\omega} \right]^2.
\end{align}
For zero temperature, an instructive interpretation of those rates in terms of a dressed energy level diagram is given in \cite{Wilson2009}.
Within the RWA, on the contrary, the corresponding rates read
\begin{equation} \label{grelRWA}
\gamma_\text{rel}^\text{RWA} = \pi G\left(\frac{\Omega_m^\text{RWA}}{2} \right) \coth \frac{\beta \hbar}{2} \Omega_m^\text{RWA} \sin^2 \Mixang{\text{RWA}},
\end{equation} 
and
\begin{equation} \label{gdephRWA}
 \gamma_\text{deph}^\text{RWA} = \half \gamma_\text{rel}^\text{RWA} + \pi N(0) \cos^2 \Mixang{\text{RWA}}.
\end{equation}
The RWA rates correspond to those of an undriven TLS \cite{Weiss2008} using the dressed energy levels and the RWA mixing angle $\Mixangle^\text{RWA}$. \newline
In Fig. \ref{fig::Rates}, we compare the rates obtained through Van Vleck perturbation theory with the RWA ones for an Ohmic spectral density, $G_\text{Ohm} (\nu) = \kappa \nu$, where $\kappa$ is the dimensionless coupling constant between TLS and bath. For both the relaxation rate, Fig. \ref{fig::Rates}(a), and  the dephasing rate, Fig. \ref{fig::Rates}(b), the RWA approach underestimates the rates. The failing of the RWA becomes especially evident in Fig. \ref{fig::Rates}(a), where a zero relaxation rate  is predicted for driving amplitudes $A$ under which $\DDelta{-m}$ vanishes. This implies in particular no relaxation at zero driving and $m \neq 0$. Again we see that the higher order matrix elements in the Floquet matrix (\ref{FloquetMatrix}) are necessary in order to correctly describe the dynamics. We find that for certain values of the driving amplitude, namely, whenever $\DDelta{-m} = 0$, $\gamma_\text{rel}^0$ vanishes and thus $\gamma_\text{rel}$ becomes minimal, a behavior which could be already predicted by inspecting formula (\ref{X-+2}) for $\X{-+}{n}$. This could be exploited experimentally to minimize relaxation.  On the other hand, for higher driving amplitudes, $\gamma_\text{deph}$ exhibits peaks at $\DDelta{-m}=0$ because of the cosine in $\gamma_\text{deph}^0$. For a high driving amplitude, our rates approach asymptotically the ones predicted by an RWA approach. In the opposite regime of small driving amplitudes, however, deviations between the RWA and Van Vleck rates occur, as matrix elements connecting the different doublets in the Floquet matrix play a more important role. Common to both approaches is that the external driving yields a reduction of the rates with increasing strength, a behavior which was already numerically predicted, e.g., in \cite{Makarov1995}.\newline
\begin{figure}
 \includegraphics[width=.45\textwidth]{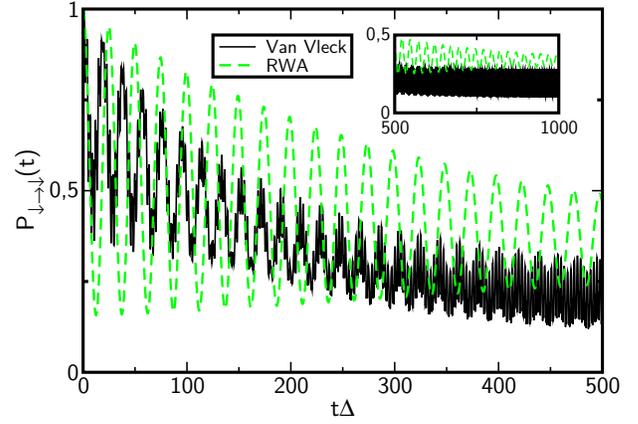}
 \caption{(Color online) Survival probability $P_{\downarrow \rightarrow \downarrow}(t)$ close to a 2-photon resonance. The parameters are $\eps /\Delta =4.1$, $\omega  /\Delta= 2.0 $, $A /\Delta=3.0$, $\kappa = 0.01$, and $ \hbar \beta \Delta= 10$. Analytical results obtained by second-order Van Vleck perturbation theory  are compared with RWA results. The inset shows the long-time dynamics and visualizes the deviation of the RWA from the Van Vleck dynamics in the asymptotic limit. \label{fig::DissDynamics}}
\end{figure}
\begin{figure}
 \includegraphics[width=.45\textwidth]{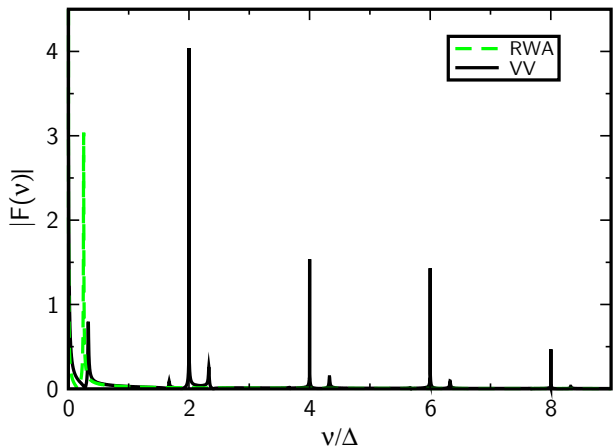}
 \caption{(Color online) Absolute value of the Fourier transform $F(\nu)$  of the survival probability  for the RWA and second-order Van Vleck perturbation theory. The parameters are the same as in Fig. \ref{fig::DissDynamics}. Next to the relaxation peak at $\nu = 0$ the RWA dynamics are governed by a single frequency $\nu =  \Omega_m^\text{RWA}$. The second-order dynamics  also exhibit the relaxation peak and a  main frequency, which, however, is shifted to $\nu =  \Omega_m^{(2)}$. Additionally, the higher harmonics of the driving can be seen in the second-order dynamics. For visualization of the $\delta$ peaks, appearing at $\nu = n \omega$, a finite width and height have been artificially introduced. Furthermore, broadened peaks appear at $\nu = n \omega \pm  \Omega_m^{(2)}$. \label{fig::FTDissDynamics}}
\end{figure}
The failure of the RWA also becomes evident in Fig. \ref{fig::DissDynamics}, where we show the dissipative dynamics obtained for an Ohmic environment. Comparing the results for $P_{\downarrow\rightarrow\downarrow}(t)$ which we obtain from second-order Van Vleck perturbation theory -- formulas  (\ref{rho--}) and (\ref{rho-+}) combined with (\ref{app::SurProb}) -- with the RWA result, we find striking differences. Considering the long-time dynamics (the inset in Fig. \ref{fig::DissDynamics}), we see that the RWA predicts quite a different asymptotic value for $P_{\downarrow\rightarrow\downarrow}(t)$. We notice further that the RWA exhibits a single oscillation frequency, which decays completely to a constant value, while within the Van Vleck solution $P_{\downarrow\rightarrow\downarrow}(t)$ oscillates for $t\rightarrow \infty$ around the equilibrium value. This latter behavior corresponds to the continuous driving of the system through the external field. It is completely missed by the RWA approach. For a further analysis of the dynamics, it is helpful to consider the Fourier transform of $P_{\downarrow\rightarrow\downarrow}(t)$, see Fig. \ref{fig::FTDissDynamics}. Both the RWA and Van Vleck dynamics exhibit a relaxation peak at $\nu = 0$ and the dressed frequency of the system at $\nu=  \Omega_m^\text{RWA}$ and $\nu =  \Omega_m^{(2)}$, respectively. Those latter peaks have a finite width due to the dephasing. Within the RWA $\Omega_m^\text{RWA}$ is the only frequency; while for second-order Van Vleck dynamics, we find additional frequencies. They result from the higher harmonics of the driving and are located at integer multiples of the driving frequency, $\nu = n \omega$, and at $\nu = n \omega \pm  \Omega_m^{(2)}$. The peaks at $\nu = n \omega$ are $\delta$ shaped as they suffer no dephasing, whereas the peaks at $\nu = n \omega \pm  \Omega_m^{(2)} $ show the broadening of the main frequency. Already in the nondissipative dynamics, Eq. (\ref{App::PVVNondiss}), we found the appearance of those multiple frequencies. They result from the beyond-RWA contributions in (\ref{PhiMinVVt}) and (\ref{PhiMaxVVt}) and reflect the external driving. Dephasing only influences the dressed frequency $\Omega_m^{(2)}$ in $\rho_{--}(t)$ and $\rho_{-+}(t)$, see (\ref{rho--}) and (\ref{rho-+}), and thus for the equilibrium state, the laser frequency at $\nu = n \omega$ is dominating. This asymptotic behavior agrees well with the findings in \cite{Grifoni1995, Grifoni1997, Grifoni1997(2)}.
\begin{figure*}
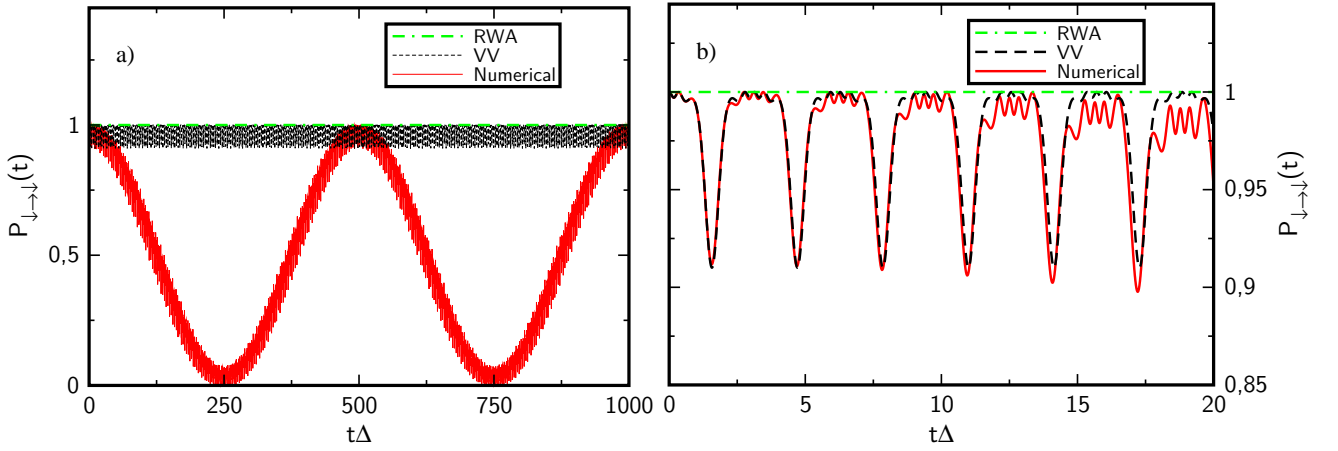

 \includegraphics[width=.48\textwidth]{Figures/Fig15_a.eps}
 \includegraphics[width=.48\textwidth]{Figures/Fig15_b.eps}
\caption{(Color online) Coherent destruction of tunneling for the nondissipative case ($\kappa=0$). The survival probability $P_{\downarrow \rightarrow \downarrow}(t)$ is shown at a 3-photon resonance. The parameters are $\eps/\Delta=6.0$, $\omega/\Delta=2.0$, and $A/\Delta=12.7603$. The Van Vleck solution  is compared with the RWA   and  a numerical diagonalization of the Floquet Hamiltonian. Within the RWA, a complete destruction of tunneling can be observed, whereas the analytic Van Vleck solution exhibits driving-induced oscillations. The numerical solution predicts, on the contrary, complete population inversion with low but nonvanishing frequency $\Omega_m$. Figure  (b) is a blowup from figure (a) for a shorter time scale. There, the numerical and Van Vleck solutions agree well, and one can nicely see the small oscillations resulting from a 3-photon absorption or emission. \label{fig::CDTNoDiss}}
\end{figure*}
\begin{figure*}
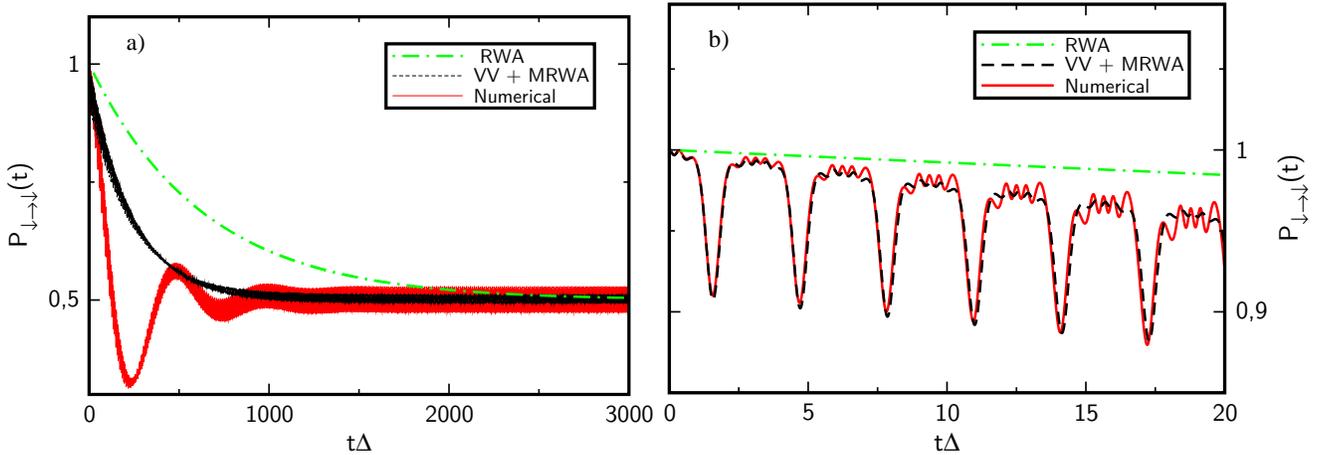

 \includegraphics[width=.48\textwidth]{Figures/Fig16_a.eps}
 \includegraphics[width=.48\textwidth]{Figures/Fig16_b.eps}
\caption{(Color online) Coherent destruction of tunneling for the dissipative case ($\kappa=0.01$, $\hbar \beta \Delta=10$). The remaining parameters are the same as in Fig. \ref{fig::CDTNoDiss}. A comparison between the numerical solution of the Floquet Hamiltonian and the full master equation (\ref{FloqMaster}), the Van Vleck combined with the MRWA approach,  and RWA results is provided. The RWA approach predicts a slower relaxation than the numerical one  and  Van Vleck solution. Deviations between the numerical and Van Vleck solutions can be seen especially in the long time limit. As in Fig. \ref{fig::CDTNoDiss}, the numerical result predicts oscillations with a nonvanishing frequency $\Omega_m$. For short times, see figure (b), the numerical and Van Vleck results agree well. \label{fig::CDTDiss}}
\end{figure*}

\subsection{Coherent destruction of tunneling} \label{sec::CDT}
 It has been found in  \cite{Grossmann1991(2),Grossmann1991(1)} for a driven double-well potential and for a driven TLS in \cite{Grossmann1992} that under certain conditions,  coherent destruction of tunneling (CDT) occurs. For a symmetric TLS ($\eps=0$) and for high enough driving frequencies ($\omega \gg  \Delta$) this phenomenon was predicted to happen approximately at the zeros of $J_0(A/\omega)$, as can also be seen from Eq. (\ref{ZeroPhotonRes}).
For a nonzero static bias and high frequencies, the necessary conditions for CDT are $\eps = m \omega$ and $J_m(A/\omega)=0$ \cite{Wang1995, Wang1996}. In this section, we compare the predictions of the RWA and Van Vleck perturbation theory against exact numerical results.\newline 
For the case of an exact $m$-photon resonance ($\eps=m\omega$) and nonvanishing $\Delta_{-m}$, the RWA mixing angle is $\Theta^\text{RWA}_m=\pi/2$, and we get from  Eq. (\ref{PSurQD}),
\begin{equation} \label{CDT3pho}
 P_{\downarrow \rightarrow \downarrow}^\text{RWA}(t) = \cos^2 \left(|J_{-m}(A/\omega)\Delta|\frac{t}{2} \right).
\end{equation}
Also from this formula, we see that CDT occurs at the zeros of $J_{-m}(A/\omega)$.
 Notice, however, that for $J_{-m}(A/\omega) =0$, Eq. (\ref{PSurQD}) predicts $P_{\downarrow \rightarrow \downarrow}^\text{RWA}(t) \equiv 1$ even for systems which are not at an $m$-photon resonance; i.e., within the RWA, the condition $\eps =m \omega$ is not necessary for CDT.\newline
 Interestingly, also  second-order Van Vleck perturbation theory predicts $\Omega_m^{(2)}=0$ for $\eps = m \omega$ and $J_{-m}(A/\omega)=0$, see Eq. (\ref{FreqVV}). However, as shown in \cite{Barata2000, Frasca2005} and discussed in Sec. \ref{sec::ValidVanVleck}, this condition holds only to second-order in $\Delta$; third-order corrections will cause $\Omega_m^{(3)}$ to be small but finite for $\eps=m \omega$ and $J_{-m}(A/\omega)=0$. Thus, instead of being localized, the dynamics will oscillate with a large period. \newline
To visualize this behavior, we examine in the following without loss of generality the case of a 3-photon resonance. We choose $\omega /\Delta = 2.0$ and $\eps /\Delta = 6.0$. Then the first zero of  $J_{-3}(A/\omega)$ occurs at $A/\Delta \approx 12.7603$.  Using those parameters in Eq. (\ref{FreqVV}), the Van Vleck oscillation frequency $\Omega_3^{(2)}$ is zero. Figure \ref{fig::CDTNoDiss} shows a comparison between the RWA and Van Vleck dynamics to second-order and an exact numerical treatment of the Floquet Hamiltonian for the above parameters. For the RWA, we see a complete destruction of tunneling because the driving-induced oscillations are not accounted for. Also, within the Van Vleck description, the  coherent oscillations are strongly suppressed; however, we notice fast oscillations because of the external driving. This becomes especially clear in Fig. \ref{fig::CDTNoDiss}(b). At $t=(2n+1) \pi/\omega$ with $n=0,1,2,3...,$ we find sharp dips. The plateaus in between show weak oscillations whose number changes with $m$. The situation changes strongly for the numerical graph: instead of a localization, a complete inversion of the population occurs; CDT seems to have vanished completely, as $\Omega_m$ is not vanishing. Considering, however, the time scale in Fig. \ref{fig::CDTNoDiss}(a), we notice that the period of $2 \pi / \Omega_m$ is rather large.  For short times, see figure  \ref{fig::CDTNoDiss}(b), also the numerical dynamics appear to be localized. Note that this observation also holds for the high-frequency case examined in \cite{Grossmann1992}: considering the dynamics at long times the localization will also be destroyed there due to higher order effects.  \newline
In Fig. \ref{fig::CDTDiss}, CDT under the influence of dissipation is examined. As in Fig. \ref{fig::CDTNoDiss}, we investigate  a 3-photon resonance with vanishing frequencies $\Omega_3^\text{RWA}$ and $\Omega_3^{(2)}$. 
We compare the dynamics obtained by a numerical solution of the Floquet-Bloch-Redfield master equation (\ref{FloqMaster}) using the exact eigenstates of the Floquet Hamiltonian, the analytical Van Vleck-MRWA approach, Eqs. (\ref{rho--}) and (\ref{rho-+}), and the RWA.
While  the Van Vleck and RWA solutions relax incoherently to a stationary state, the numerical solution exhibits two full oscillations with $\Omega_m$. As in the nondissipative case, the exact oscillation frequency $\Omega_m$ is nonzero. For stronger damping those slow oscillations disappear. Both the numerical and Van Vleck oscillations show fast driving-induced oscillations which survive also in the stationary state. While for short time scale, Fig. \ref{fig::CDTDiss}(b), both approaches agree quite well, one finds that in the long time limit the amplitude of the fast oscillations predicted by the analytical solution is smaller than the exact numerical one. Compared to the RWA solution, where the equilibrium value is reached after longer time and the fast oscillations are averaged out,  Van Vleck perturbation theory is clearly an improvement.
\begin{figure*}
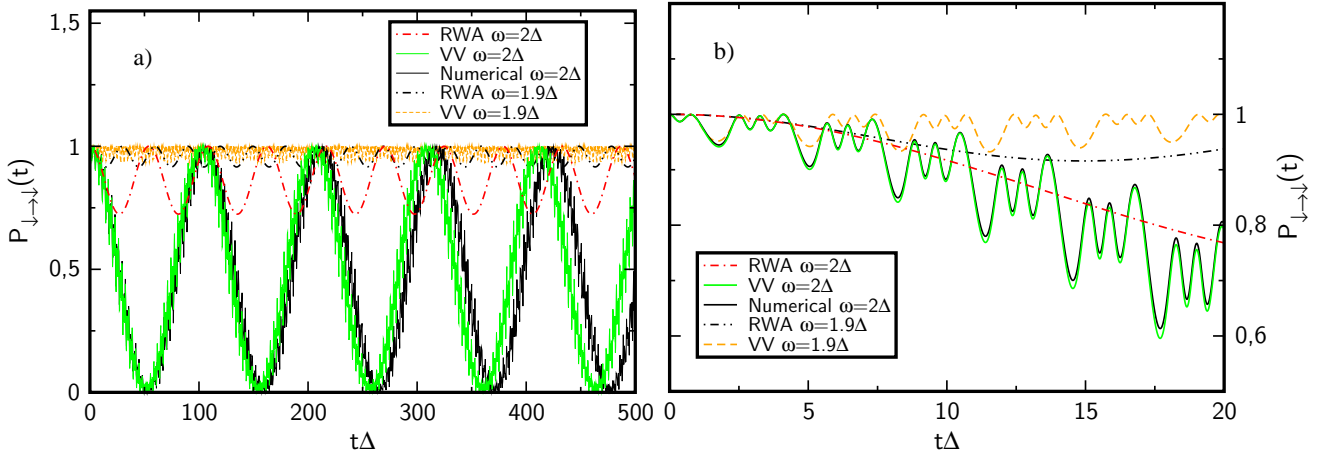

 \includegraphics[width=.48\textwidth]{Figures/Fig17_a.eps}
 \includegraphics[width=.48\textwidth]{Figures/Fig17_b.eps}
 \caption{(Color online) Driving-induced tunneling oscillations for the nondissipative case ($\kappa=0$). The survival probability $P_{\downarrow \rightarrow \downarrow}(t)$ is shown for $A/\Delta=3.0$, $\omega/\Delta=2.0$, and $\eps/\Delta=5.9011$ (exact 3-photon resonance). Three approaches are compared: a complete numerical solution of the Floquet Hamiltonian,  the second-order Van Vleck approach,  and the RWA approach. For the first two approaches, complete population inversion is predicted, and  for the Van Vleck dynamics we find the main oscillation frequency $\Omega_3^{(2)}=\Delta |J_{-3}(A/\omega)|$. Besides, the numerical and Van Vleck approaches exhibit small driving-induced oscillations, see especially figure (b).  The RWA predicts a strongly shifted oscillation frequency. Moreover, population inversion is incomplete. For further comparison, the RWA  and the Van Vleck approaches  are shown for a slightly shifted external frequency, $\omega /\Delta = 1.9$. The dynamics in this case are almost completely localized. \label{fig::DITONoDiss}}
\end{figure*}
\begin{figure*}
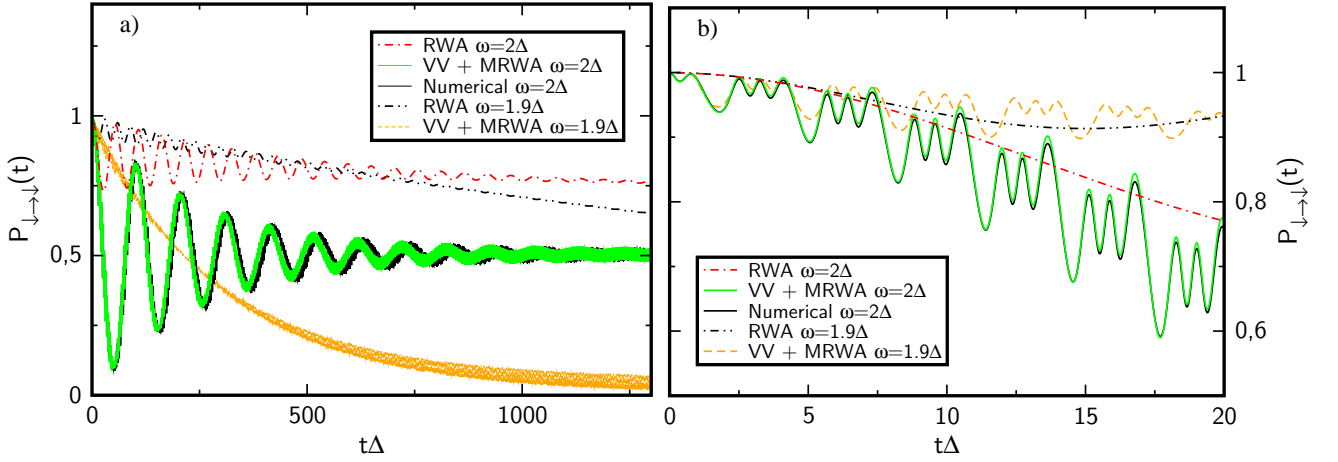

 \includegraphics[width=.48\textwidth]{Figures/Fig18_a.eps}
 \includegraphics[width=.48\textwidth]{Figures/Fig18_b.eps}
 \caption{(Color online) Driving-induced tunneling oscillations for the dissipative case ($\kappa=0.01$ and $\hbar \beta \Delta=10$). Remaining parameters are the same as in Fig. \ref{fig::DITONoDiss}. The numerical solution of the master equation (\ref{FloqMaster}), the analytical Van Vleck-MRWA solution, and the RWA solution for $\omega/\Delta=2.0$ are compared. Good agreement between the numerical and the Van Vleck solutions can be observed on both long (a) and short  (b) time scales.  Within the RWA, not only are the main frequency and  amplitude  changed and the driving-induced oscillations missed, but also the equilibrium value lies far above the Van Vleck prediction. The RWA  and the  Van Vleck-MRWA  solution for $\omega/\Delta =1.9$  show an almost incoherent decay, and their long time limits differ strongly from the corresponding ones for $\omega/\Delta=2.0$. \label{fig::DITODiss}}
\end{figure*}

\subsection{Driving-induced tunneling oscillations} \label{sec::DITO}
An effect contrary to the CDT are driving-induced tunneling oscillations (DITO).  It has been predicted in \cite{Hartmann1998, Hartmann2000, Goychuk2005} and experimentally shown in \cite{Nakamura2001} that for a high static energy bias, $\eps \gg \Delta$, and for high driving frequency, $\omega \gg \Delta$, coherent oscillations with frequency $\Delta |J_{-m}(A/\omega)|$ and large amplitude are induced if $\eps \approx m \omega$.  The DITO are often also named Rabi oscillations even though in the original problem of Rabi \cite{Rabi1937} a circularly polarized driving field couples to the TLS. As a consequence, the obtained frequency of the oscillations is linear in $A$. \newline
In this section, we are going to investigate the effect in the regime of moderate energy bias and driving frequency.    
First, we examine again the nondissipative case ($\kappa=0$), see Fig. \ref{fig::DITONoDiss}. As parameters, we choose a moderate driving amplitude and frequency: $A/\Delta = 3.0$ and $\omega/\Delta = 2.0$. For an exact 3-photon resonance, condition (\ref{ResCondVV}) must be fulfilled and thus $\eps/\Delta \approx 5.9011$. Notice that the RWA resonance condition (\ref{ResCond}), $\eps = 6 \omega$, is only valid in the case of high frequencies $\omega \gg \Delta$. With condition (\ref{ResCondVV}) used, the Van Vleck approach results in the oscillation frequency $\Omega_3^{(2)} =  \Delta |J_{-3}(A/\omega)|$, and for times $t=(2n+1)\pi/\Omega_3^{(2)}$ with $n=0,1,2,3...,$ one finds complete population inversion, see the Van Vleck graph in Fig. \ref{fig::DITONoDiss}(a). Furthermore, in Fig. \ref{fig::DITONoDiss}(b), one can nicely see the modifications resulting  from the external driving: three small oscillations corresponding to a 3-photon resonance. The exact numerical solution shows a slightly shifted main oscillation frequency $\Omega_3$. The RWA approach  exhibits the oscillation frequency $\Omega_3^\text{RWA}$, which is strongly out of phase compared to the numerical and Van Vleck one, and also has a smaller amplitude, so that a complete population inversion is not reached. When changing the driving frequency to slightly out of resonance, the driving-induced tunneling oscillations are strongly suppressed, and the system is almost completely localized in the initial state.  This behavior originates -- contrary to the CDT -- not in a zero oscillation frequency $\Omega_3^{(2)}$ but rather in a vanishing amplitude of the $\Omega_3^{(2)}$ oscillation. Also the RWA at $\omega =1.9 \Delta$ is suppressed.\newline
In Fig. \ref{fig::DITODiss}, we consider the influence of the environment. At exact resonance, we observe within the numerical solution and the Van Vleck-MRWA approach coherent oscillations decaying to a stationary equilibrium value. Before reaching the equilibrium value, the dynamics are dominated by the frequency $\Omega_3$, while in the long time limit the coherent oscillations die out; faster ones with the driving frequency $\omega$ and its higher harmonics around a static equilibrium value are found. The agreement between the numerical and analytical calculations is quite good. 
Also, in the RWA approach,  the coherent oscillation of frequency $\Omega_3^\text{RWA}$ dies out to a stationary state. However, apart from the  frequency shift already observed in the nondissipative case and the smaller amplitude, the equilibrium value also differs strongly from the one obtained within the Van Vleck solution. Furthermore, since fast oscillations are completely neglected, the stationary state is constant. Considering the Van Vleck solution for a slightly shifted driving frequency, $\omega/\Delta = 1.9$,  we notice an almost incoherent decay to an equilibrium value which is much lower than the one of the dynamics with $\omega/\Delta=2.0$. Thus, dissipation leads here to an almost complete inversion of the population.\newline
 We observe that our analytical methods are also able to recover the findings for the population difference in Chapter 3.2 of \cite{Goychuk2005} in the high-frequency limit ($\omega\gg\Delta$) and even can reproduce the small modulations which are found there by a numerical treatment of the dynamics. Furthermore, we are able to go beyond the assumption of a high driving frequency.

\section{Conclusions}\label{sec::Conclusion}
In conclusion, we discussed the dynamics of the spin-boson system exposed to an external ac driving. Assuming weak coupling between TLS and bath, we arrived at a closed analytical expression for the time evolution of the system. Our results are at the same time valid for the  whole range of the driving amplitude $A$ and for  moderate to high driving frequencies $\omega$, see discussion in Sec. \ref{sec::ValidVanVleck}. In contrast to the NIBA, we are able to treat both an unbiased and a biased TLS for low temperatures and weak damping. Indeed, besides the Born-Markov approximation, the only further simplifications we used solving the time-dependent Hamiltonian are the moderate rotating-wave approximation in Sec. \ref{sec::MRWA} and the expansion in the dressed tunneling matrix element $\DDelta{n}=J_n(A/\omega) \Delta$, with $\Delta$ the bare tunneling coupling and $J_n$ the $n$th-order Bessel function, using Van Vleck perturbation theory. In the vicinity of an $m$-photon resonance, the latter is also justified, as shown in Sec. \ref{sec::ValidVanVleck}, for moderate driving frequencies as long as condition (\ref{VanVleckCond}) is valid.
We found corrections to the renormalized Rabi frequency $\Omega_m^\text{RWA}$ [Eq. (\ref{FreqRWA})]  also leading to a shift of the resonance condition for an $m$-photon resonance [Eq. (\ref{ResCondVV})]. The so-calculated quasienergy spectrum is in very good agreement with results found by a numerical diagonalization of the Floquet Hamiltonian for all values of the static bias $\eps$. Upon investigation of the survival probability $P_{\downarrow \rightarrow \downarrow}(t)$, we could recover the shifted oscillation frequency reported already in \cite{Son2009}. We included also the second-order modifications to the Floquet states in our calculation, which account for the higher harmonics induced through the external driving and lead to fast oscillations in $P_{\downarrow \rightarrow \downarrow}(t)$, see Figs. \ref{fig::PNonDiss}, \ref{fig::DissDynamics}, and \ref{fig::FTDissDynamics}.\newline
By adding a thermal bath to the TLS, we examined in Sec.  \ref{sec::DissSys} the dissipative dynamics of the system. 
In Sec. \ref{sec::PosMat}, we visualized the good agreement between our analytical formulas for the position matrix elements and a numerical calculation even for low driving amplitudes. This turned out to be essential to arriving at a physically realistic result for the relaxation and dephasing rates given in Sec. \ref{sec::MRWA}. Comparing RWA to Van Vleck results, we found strong deviations and even unphysical predictions for the former one at low driving amplitudes.
 We remark that our rates agree very well with the zero-temperature results derived recently in \cite{Wilson2009} via the dressed state approach. In this work, a charge qubit is strongly driven by a microwave field and connected to a dc SQUID. The higher order corrections to the rates prove to be essential to  correctly reflecting the physical findings in this experiment. From this we are encouraged that our results provide a realistic picture of relaxation and dephasing processes in a driven two-level system and, due to the generality of the model, are of interest to a wide range of physical applications.\newline
In Sec. \ref{sec::CDT}, we performed a detailed analysis of the TLS at an exact 3-photon resonance and for a vanishing third-order Bessel function, which is known to lead to coherent destruction of tunneling in the high-frequency limit. For moderate driving frequencies, we found second-order modifications to the  RWA solution. While the latter predicts a complete localization of the TLS in the initial state, the Van Vleck solution shows that driving-induced oscillations survive.  Furthermore, for the dissipative case  we found an incoherent decay to a quasistationary value.\newline
In Sec. \ref{sec::DITO}, we examined an effect opposite to the coherent destruction of tunneling: for an appropriately chosen driving amplitude, coherent tunneling oscillations with frequency $ \Omega_m$, Eq. (\ref{FreqVV}), can be observed at an $m$-photon resonance. By slightly changing the driving frequency out of resonance, these oscillations are almost completely suppressed and the system shows an incoherent behavior.

\begin{acknowledgments}
 We acknowledge financial support under DFG Program SFB631. Further we would like to thank Chris M. Wilson for helpful discussions and Marco Frasca for constructive remarks.
\end{acknowledgments}

\appendix
\section{Floquet theory}\label{app::FloquetTheory}
 The Floquet theorem states that the Schr\"odinger equation
\begin{equation} \label{SGLH}
   \rmi \hbar \partial_t \ket{\psi(t)} = H(t) \ket{\psi(t)}
 \end{equation} 
for a Hamiltonian being periodic in time is solved by
\begin{equation} \label{SolutionH}
  \ket{\psi_\alpha (t)} = \ket{u_\alpha(t)} \rme^{-\rmi \varepsilon_\alpha t /\hbar},
\end{equation} 
where $\ket{u_\alpha(t)} = \ket{u_\alpha(t+T)}$ and we assume that the oscillation period of $H(t)$ is $T \equiv 2\pi/\omega$. The quasienergies $\varepsilon_\alpha$ are obtained as eigenvalues of the Floquet Hamiltonian
\begin{equation}
 \fh  \ket{u_\alpha(t)} =[H(t) - \rmi \hbar \partial_t] \ket{u_\alpha(t)} = \varepsilon_\alpha \ket{u_\alpha(t)}. 
\end{equation} 
Note that $\ket{u_{\alpha, n} (t)} \equiv \exp(-\rmi n \omega t) \ket{u_{\alpha}(t)}$ yields a solution  of (\ref{SGLH}) physically identical  to (\ref{SolutionH}) but with the shifted quasienergy $\varepsilon_{\alpha,n} \equiv \varepsilon_\alpha - \hbar n \omega$. Furthermore, $\eps_{\alpha} = \eps_{\alpha,0}$ and $\ket{u_\alpha(t)} = \ket{u_{\alpha,0}(t)}$. Thus, it will be sufficient just to examine the set of eigenvalues $\{ \varepsilon_{\alpha,n} \}$ with $- \hbar \omega/2 \leq \varepsilon_{\alpha,n} < \hbar \omega/2$.\newline
We introduce the Hilbert space $\mathcal{T}$ of the $T$-periodic functions, with the inner product defined as
\begin{equation}
  (f,g) = \frac{1}{T} \int_0^T dt \, f^*(t) g(t).
\end{equation} 
The functions $\varphi_{n}(t) = \exp(-\rmi n \omega t)$ build an orthonormal and complete basis set of $\mathcal{T}$ \cite{Simmons1963}, where we further define for a basis-independent notation the state vectors $\ketT{n}$ with $\varphi_{n}(t) =  \braketT{t}{n}$.
The scalar product in the extended Hilbert space $\fh \otimes \mathcal{T}$ of the Floquet Hamiltonian is defined as 
\begin{equation} \label{InternalProduct}
  \braketfl{\cdot}{\cdot} = \frac{1}{T} \int_{0}^T \, dt \braket{\cdot}{\cdot}.
\end{equation} 
Considering a $T$-periodic state vector $\ket{u_{\alpha,n}(t)}$ living in the spatial Hilbert space $\mathcal{H}$, we can write it in a Fourier series and thus expand it in basis functions of $\mathcal{T}$:
\begin{equation}
  \ket{u_{\alpha,n}(t)} = \rme^{-\rmi n \omega  t} \ket{u_\alpha (t)}  = \sum_l \rme^{-\rmi l \omega t} \ket{u_\alpha^{(n-l)}},
\end{equation} 
where $\ket{u_\alpha^{(k)}}$ are the time independent Fourier coefficients. In the composite Hilbert space $\fh \otimes \mathcal{T}$, we define the state
\begin{equation} \label{Composite state}
  \ketfl{u_{\alpha,n}} \equiv \sum_l  \ket{u_\alpha^{(n-l)}}\otimes   \ketT{l}.
\end{equation}
Through the expansion of the Hilbert space, we can now treat the time-dependent problem (\ref{SGLH}) like a time independent one.

\section{Van Vleck perturbation theory} \label{app::VanVleck}
Here, we give the matrix elements of the effective Hamiltonian $\fh_\text{eff}$ and the transformation matrix $\exp(\pm \rmi S)= \mathds{1} \pm \rmi \str{1} \pm \rmi \str{2} + \half \rmi \str{1} \rmi \str{1}$ to second-order in $\Delta$ expressed in the eigenstates (\ref{BasisUHT}) of the unperturbed Hamiltonian. For first order in $\Delta$, the Hamiltonian has the same shape as within the RWA. Its elements are \cite{CohenTannoudji2004, Hausinger2008}
\begin{align}
 \brafl{u_{\uparrow/\downarrow,n}^0} \fh_\text{eff}^{(1)}\ketfl{u_{\uparrow/\downarrow,l}^0} &= \hbar \eps_{n}^0 \delta_{n,l}, \\
 \brafl{u_{\uparrow,n}^0} \fh^{(1)}_\text{eff} \ketfl{u_{\downarrow,l}^0} &=- \hh \DDelta{n-l} \delta_{n-l,m}.
\end{align} 
The elements of the transformation matrix are
\begin{align}
  \ulfMin{n} \rmi \str{1} \uPlus{l} &=  \half \frac{\DDelta{n-l}}{\eps+\omega(n-l)} (1-\delta_{l-n,m}), \label{S11}\\
  \ulfPlus{l} \rmi \str{1} \uMin{n} &=  - \half \frac{\DDelta{n-l}}{\eps+\omega(n-l)} (1-\delta_{l-n,m}). \label{S12}
\end{align}
The Kronecker $\delta$ comes from the fact that $\rmi \str{1}$ vanishes between almost degenerate states by construction.
For the second-order elements, we find
\begin{align}
  \brafl{u_{\uparrow/\downarrow,n}^0} \fh_\text{eff}^{(2)}\ketfl{u_{\uparrow/\downarrow,l}^0} &=   \mp \frac{\hbar}{4} \sum_{l \neq -m} \frac{|\DDelta{l}|^2}{\eps + l \omega} , \\
   \ulfMin{n} \fh_{\rm eff}^{(2)} \uPlus{n+m} &=  \ulfPlus{n+m} \fh_{\rm eff}^{(2)} \uMin{n} = 0.
\end{align}
The expression for the transformation matrix already becomes more evolved:
\begin{widetext}
\begin{align}
  \ulfMin{n} \rmi \str{2} \uMin{j} & = \frac{1}{4 (n-j) \omega} \biggl\{ \mathop{\sum_{k \neq n+m}}_{k \neq j+m} \frac{\DDelta{n-k} \DDelta{j-k}}{2} \biggl[ \frac{1}{ \eps + (n-k) \omega } + \frac{1}{\eps + (j-k) \omega} \biggr] \nonumber \\
               & + \frac{\DDelta{n-j-m} \DDelta{-m}}{\eps + (n-j-m) \omega} + \frac{\DDelta{j-n-m} \DDelta{-m}}{\eps + (j-n-m) \omega} \biggl\}(1-\delta_{j,n}),
\end{align}
\begin{align}
  \ulfPlus{n} \rmi \str{2} \uPlus{j} & = \frac{1}{4 (n-j) \omega} \biggl\{ \mathop{\sum_{k \neq n-m}}_{k \neq j-m} \frac{\DDelta{k-n} \DDelta{k-j}}{2} \biggl[ \frac{1}{ -\eps + (n-k) \omega } + \frac{1}{-\eps + (j-k) \omega} \biggr] \nonumber \\
               & + \frac{\DDelta{j-m-n} \DDelta{-m}}{-\eps + (n-j+m) \omega} + \frac{\DDelta{n-m-j} \DDelta{-m}}{-\eps + (j-n+m) \omega} \biggl\}(1-\delta_{j,n}),
\end{align}
\begin{equation}
  \ulfMin{n} \rmi \str{2} \uPlus{j} = \ulfPlus{j} \rmi \str{2} \uMin{n} =0.
\end{equation} 
By applying the transformation now on the eigenstates of the effective Hamiltonian $\ketfl{\Phi_{\mp,n}^\text{eff}}$, see Sec. \ref{sec::VanVleck}, we get the eigenstates of the Floquet Hamiltonian $\fh_\text{TLS}$ to first order in $\Delta$,
\begin{align}
 \PflMin{n}{(1)} = & \PflMin{n}{\text{eff}} + \half  \sum_{j\neq -m} \frac{\DDelta{j}}{\eps + j \omega} \left\{  \sign{\Dm}  \cosmixa\uMin{j+n+m} -  \sinmixa   \uPlus{n-j} \right\}, \label{app::FiMin1}
\end{align}
\begin{align}
 \PflPlus{n}{(1)} = &\PflPlus{n}{\text{eff}} + \half \sum_{j \neq -m} \frac{\DDelta{j}}{\eps + j \omega} \left\{ \sign{\Dm}  \sinmixa   \uMin{j+n}  +  \cosmixa   \uPlus{n-m-j} \right\}. \label{app::FiPlus1}
\end{align}
And to second-order,
\begin{align} 
  & \PflMin{n}{(2)}= \PflMin{n}{(1)} + \sum_{j\neq 0} \left\{ \sinmixa \uMin{j+n} + \sign{\Dm} \cosmixa \uPlus{-j+n+m}  \right\}  \nonumber \\
      & \phantom{\mathrel{+}} \times \left\{  \frac{\DDelta{-m}}{4 j\omega} \left[ \frac{\DDelta{j-m}}{\eps + (j-m)\omega} + \frac{\DDelta{-j-m}}{\eps - (j+m)\omega}  \right] +  \mathop{\sum_{p\neq -j-m}}_{p \neq -m}  \frac{1}{4 j \omega} \frac{\DDelta{j+p} \DDelta{p}}{2} \left[ \frac{1}{\eps + (j+p) \omega} + \frac{1}{\eps +p\omega} \right] \right\} \nonumber\\
     & + \frac{1}{8} \sum_{k\neq -m} \sum_{j\neq -m} \frac{\DDelta{k} \DDelta{j}}{(\eps+k\omega)(\eps+j\omega)} \left\{ \sinmixa \uMin{k+n-j} + \sign{\Dm} \cosmixa \uPlus{k+n+m-j}  \right\} , \label{app::FiMin2}
\end{align}
\begin{align}  
& \PflPlus{n}{(2)}= \PflPlus{n}{(1)} - \sum_{j\neq 0} \left\{ \cosmixa \uMin{j+n-m} - \sign{\Dm} \sinmixa \uPlus{-j+n} \right\} \nonumber \\
      & \phantom{\mathrel{+}} \times \left\{  \frac{1}{4 j\omega} \left[ \frac{\DDelta{j-m} \DDelta{-m}}{\eps + (j-m)\omega} + \frac{\DDelta{-j-m} \DDelta{-m}}{\eps - (j+m)\omega}  \right] +  \mathop{\sum_{p\neq -j-m}}_{p \neq -m}  \frac{1}{4 j \omega} \frac{\DDelta{j+p} \DDelta{p}}{2} \left[ \frac{1}{\eps + (j+p) \omega} + \frac{1}{\eps +p\omega} \right] \right\} \nonumber \\
     & - \frac{1}{8} \sum_{k\neq -m} \sum_{j\neq -m} \frac{\DDelta{k} \DDelta{j}}{(\eps+k\omega)(\eps+j\omega)} \left\{ \cosmixa \uMin{k+n-j-m} - \sign{\Dm} \sinmixa \uPlus{j+n-k}  \right\}. \label{app:FiPlus2}
\end{align}

\end{widetext}

\section{Calculation of the dynamics} \label{app::dynamics}
To calculate the survival probability of the system, $P_{\gamma \rightarrow \gamma}(t)$, where $\gamma = \uparrow, \downarrow$, we start with the density matrix $\rho(t)$ of the TLS, fulfilling the condition that $\rho(0) = \ket{\gamma}\bra{\gamma}$. By diagonalization of the Floquet matrix (\ref{FloquetMatrix}) or by solving the master equation (\ref{FloqMaster}), we obtain the density matrix in energy basis with the matrix elements
\begin{equation}
  \rho_{\alpha \beta}(t) = \bra{\Phi_\alpha(t)} \rho(t) \ket{\Phi_\beta (t)} \qquad \alpha, \beta = \pm. 
\end{equation} 
Using that $\rho_{--}(t) + \rho_{++}(t) =1$ and $\rho_{-+} (t) = \rho_{+-}^*(t)$, we get
\begin{align} \label{app::SurProb}
   &P_{\gamma \rightarrow \gamma} (t) = 2 \text{Re} \left\{ \braket{\gamma}{\Phi_-(t)}   \braket{\Phi_+(t)}{\gamma} \rho_{-+}(t) \right\} \nonumber \\
     & + |\braket{\gamma}{\Phi_+(t)}|^2 + \left( |\braket{\gamma}{\Phi_-(t)}|^2 - |\braket{\gamma}{\Phi_+(t)}|^2  \right) \rho_{--}(t).
\end{align} 
The corresponding transition probability is just $P_{\gamma \rightarrow \delta}(t) = 1- P_{\gamma \rightarrow \gamma} (t) $, where $\delta \neq \gamma$. From (\ref{app::SurProb}), we see that we have to calculate $\braket{\gamma}{\Phi_\alpha(t)}$. We use the periodicity in time and express it in a Fourier series:
\begin{equation}
  \braket{\gamma}{\Phi_\alpha(t)} = \sum_k \braket{\gamma}{\Phi_\alpha^{(k)}} \exp(\rmi k \omega t),
\end{equation} 
 with
\begin{align}
  \braket{\gamma}{\Phi_\alpha^{(k)}} &= \frac{1}{T} \int_0^T \, dt \rme^{-\rmi k \omega t} \braket{\gamma}{\Phi_\alpha(t)} \nonumber \\
                               &= \frac{1}{T} \int_0^T \, dt  \braket{\gamma}{\Phi_{\alpha,k}(t)} = \braketfl{\gamma}{\Phi_{\alpha,k}},
\end{align} 
where we used in the last step the definition for the inner product of the extended Hilbert space, Eq. (\ref{InternalProduct}), and defined $\ketfl{\gamma}\equiv \ket{\gamma} \ketT{0}$.  By this we establish a connection between the Floquet states and the time-dependent Hilbert state:
\begin{equation} \label{app::Fit}
 \braket{\gamma}{\Phi_\alpha(t)} = \sum_k \braketfl{\gamma}{\Phi_{\alpha,k}} \rme^{\rmi k \omega t}.
\end{equation} 

\subsection{Survival probability in the nondissipative case}
For the Hamiltonian of the nondissipative TLS, Eq. (\ref{FreeHamil}), the master equation is simply
\begin{equation}
  \dot \rho_{\alpha\beta} (t) = - \rmi (\eps_{\alpha} - \eps_{\beta}) \rho_{\alpha \beta}(t),
\end{equation} 
so that
\begin{equation}
 \rho_{--}(t) = \rho_{--}(0),
\end{equation} 
  and
\begin{equation}
 \rho_{-+}(t) = \rho_{-+}(0) \exp \left[\rmi \left(m \omega +  \Omega_m^{(2)} \right)t \right],
\end{equation} 
where we used the general expression for the quasienergies at an $m$-photon resonance found in Sec. \ref{sec::NonDiss}. The starting conditions are calculated through 
\begin{equation} \label{StartCond}
\rho_{\alpha \beta}(0) = \braket{\Phi_{\alpha}(0)}{\gamma}\braket{\gamma}{\Phi_\beta(0)}.
\end{equation} 
Combing this, one gets 
\begin{align} \label{app::Pdowndown}
    P_{\gamma \rightarrow \gamma} (t) &=  \left( |\braket{\gamma}{\Phi_-(t)}|^2 - |\braket{\gamma}{\Phi_+(t)}|^2  \right) |\braket{\gamma}{\Phi_-(0)}|^2 \nonumber \\
      & + |\braket{\gamma}{\Phi_+(t)}|^2 + 2 \text{Re} \biggl\{ \braket{\gamma}{\Phi_-(t)}   \braket{\Phi_+(t)}{\gamma}  \nonumber \\
  & \phantom{\mathrel{=}} \times \braket{\Phi_-(0)}{\gamma} \braket{\gamma}{\Phi_+(0)} \rme^{\rmi \left(m \omega + \Omega_m^{(2)} \right) t}   \biggr\}.
\end{align}\newline

\subsubsection{RWA survival probability}
Using in this general expression the eigenstates (\ref{FiMinQD}) and (\ref{FiPlusQD}), we arrive at the survival probability in the RWA,
\begin{equation} 
  P_{\downarrow \rightarrow \downarrow}^\text{RWA} (t) = \cos^2 \left( \Omega^\text{RWA}_m \frac{t}{2} \right) + \cos^2 \Mixang{\text{RWA}} \sin^2 \left( \Omega^\text{RWA}_m \frac{t}{2}\right). 
\end{equation}\newline

\subsubsection{Van Vleck survival probability}
To get the survival probability to second-order in $\Delta$, we use (\ref{app::FiMin1}) -- (\ref{app:FiPlus2}) in (\ref{app::Fit}) and obtain
\begin{widetext}
\begin{align}
  \braket{\downarrow}{\Phi_{-}^{(2)}(t)} &= \exp \left( -\rmi \frac{A}{2 \omega} \sin \omega t \right) \biggl\{ -\sign{\DDelta{-m}} \cosmixa \rme^{-\rmi m \omega t} - \half \sinmixa A(t) \nonumber  \\
              & + \sign{\DDelta{-m}}  \cosmixa  \rme^{-\rmi m \omega t} \left[ B(t) + C(t)  \right]\biggr\}, \label{PhiMinVVt}
\end{align}
\begin{align}
  \braket{\downarrow}{\Phi_{+}^{(2)}(t)} &= \exp \left( -\rmi \frac{A}{2 \omega} \sin \omega t \right) \biggl\{ -\sign{\DDelta{-m}} \sinmixa  + \half \cosmixa \rme^{\rmi m \omega t} A(t) \nonumber  \\
              & + \sign{\DDelta{-m}}  \sinmixa   \left[ B(t) - C^*(t)  \right]\biggr\},\label{PhiMaxVVt}
\end{align}
where we defined
\begin{align}
  A(t) &\equiv \sum_{n\neq-m} \rme^{\rmi n \omega t} \frac{\DDelta{n}}{\eps + n \omega},
\end{align}
\begin{align}
    B(t) &= \sum_{n\neq0} \rme^{\rmi n \omega t} \frac{1}{4 n \omega} \left[ \frac{\DDelta{n-m} \DDelta{-m}}{\eps + (n-m)\omega} + \frac{\DDelta{-m-n} \DDelta{-m}}{\eps -(n+m) \omega} \right] + \frac{1}{8} |A(t)|^2,
\end{align}
\begin{align}
  C(t) &= \sum_{n\neq 0} \mathop{\sum_{p\neq -m}}_{p\neq - n-m} \frac{\DDelta{p} \DDelta{p+n}}{8 n \omega} \rme^{\rmi n \omega t} \left[ \frac{1}{\eps + p \omega} + \frac{1}{\eps + (p+n) \omega}  \right]
\end{align}
Using those expressions in (\ref{app::Pdowndown}), we obtain the survival probability 
\begin{equation} \label{App::PVVNondiss}
 P_{\downarrow \rightarrow \downarrow}(t) = P_{\downarrow \rightarrow \downarrow}^{\text{RWA}^\prime}(t) + P_{\downarrow \rightarrow \downarrow}^{(1)}(t) +  P_{\downarrow \rightarrow \downarrow}^{(2)}(t).
\end{equation} 
We distinguish three different parts. The first one corresponds to the averaged second-order Van Vleck approach:
\begin{equation}
 P_{\downarrow \rightarrow \downarrow}^{\text{RWA}^\prime}(t) = \cos^2 \left(\Omega_m^{(2)} \frac{t}{2}\right) + \cos^2 \Mixangle \sin^2 \left( \Omega_m^{(2)} \frac{t}{2}\right). 
\end{equation}  
Then we have additional contributions from Van Vleck perturbation theory to first order in $\Delta$:
\begin{align}
  & P_{\downarrow \rightarrow \downarrow}^{(1)}(t) = -\half \sign{\Delta_m} \sin \Mixangle \, \sin  \Omega_m^{(2)} t \sum_{n\neq-m} \frac{\DDelta{n}}{\eps+ n \omega} \sin [(n+m) \omega t] \nonumber \\
          &+ \sign{\Delta_m} \sin 2\Mixangle \, \sin^2  \left(\Omega_m^{(2)} \frac{t}{2} \right) \left[ \sum_{n \neq -m} \half \frac{\DDelta{n}}{\eps+ n \omega} \cos (n+m) \omega t  + \half A(0)   \right].
\end{align}
And finally the second-order part:
\begin{align}
 & P_{\downarrow \rightarrow \downarrow}^{(2)}(t) = -\sign{\Delta_m} \sin 2\Mixangle \, \sin^2 \left( \Omega_m^{(2)} \frac{t}{2} \right) \sqrt{\frac{1}{4} A(0)^2-2[B(0)+C(0) \cos \Mixangle]} \nonumber \\
& + \sin^2 \Mixangle \sin^2 \left( \Omega_m^{(2)} \frac{t}{2} \right) \left[ \frac{1}{4} \left( \sum_{n \neq -m}  \frac{\DDelta{n}}{\eps+ n \omega} \sin (n+m) \omega t \right)^2 + \frac{1}{4} \left( \sum_{n \neq -m}  \frac{\DDelta{n}}{\eps+ n \omega} \cos (n+m) \omega t \right)^2 \right. \nonumber \\
           &+ \left. \left( \half A(0)- \sqrt{\frac{1}{4} A(0)^2-2[B(0)+C(0) \cos \Mixangle]}  \right)^2 \right]+ \left( \half A(0)- \sqrt{\frac{1}{4} A(0)^2-2[B(0)+C(0) \cos \Mixangle]}  \right) \nonumber \\
           &\sum_{n \neq -m} \frac{\DDelta{n}}{\eps +n \omega}  \left[ \cos(n+m) \omega t \cos^2 \left( \Omega_m^{(2)} \frac{t}{2} \right) +\cos \Mixangle \sin(n+m) \omega t \sin  \Omega_m^{(2)} t - \cos(n+m)\omega t \cos^2 \Mixangle \sin^2 \left( \Omega_m^{(2)} \frac{t}{2}\right)  \right] \nonumber \\ 
            &- \left[\cos^2 \left(\Omega_m^{(2)} \frac{t}{2}\right) - \cos^2 \Mixangle \sin^2 \left(\Omega_m^{(2)} \frac{t}{2}\right) \right] \sum_{n \neq 0} \frac{1}{2 n \omega} \left[ \frac{\DDelta{n-m} \DDelta{-m}}{\eps + (n-m) \omega} + \frac{\DDelta{-m-n} \DDelta{-m}}{\eps -(n+m) \omega}   \right] \cos n \omega t \nonumber\\
           &- \frac{1}{4} \left[ \cos^2 \left(\Omega_m^{(2)} \frac{t}{2}\right)   + \cos^2 \Mixangle \sin^2 \left(\Omega_m^{(2)} \frac{t}{2}\right) \right] \sum_{j \neq -m} \sum_{k\neq -m} \frac{\DDelta{j} \DDelta{k}}{(\eps + k \omega) (\eps + j \omega)} \cos(j-k) \omega t - f \cos \Mixangle\nonumber \\
           &+ \sum_{n \neq 0} \mathop{\sum_{p \neq -m}}_{p \neq -n - m} \frac{\DDelta{p} \DDelta{p+n}}{4 n \omega} \left[ \frac{1}{\eps + p \omega} + \frac{1}{\eps+(p+n) \omega}  \right] \cos \Mixangle \cos n \omega t -2 d \left[ \cos^2 \left(\Omega_m^{(2)} \frac{t}{2}\right) + \cos \Mixangle \sin^2 \left(\Omega_m^{(2)} \frac{t}{2}\right) \right].
\end{align}
Here, for $\eps \neq m \omega$, we have
\begin{align}
  d &= \frac{\Delta^2}{\frac{1}{4} a^2 - 2 b - 2 c \cos \Mixang{2}} \left[\frac{1}{16} a - \frac{3}{4} b a^2 + 2 b + 2 c^2 - 2 c (\frac{1}{2}a^2- 3 b) \cos \Mixang{2} + 2 c^2 \cos 2 \Mixang{2} \right. \nonumber\\
       &- \left. \frac{1}{2} a \left( \frac{1}{4} a^2 - 2 b - c -3 c \cos \Mixang{2}\right) \sqrt{\frac{1}{4} a^2 - 2 b - c - 3 c \cos \Mixang{2}} \right]
\end{align}
and
\begin{align}
  f &= \frac{\Delta^2}{\frac{1}{4} a^2 - 2 b - 2 c \cos \Mixang{2}} \left[\frac{1}{4} c a^2 - 2b c - 2c^2 \cos \Mixang{2} \right]
\end{align}
with $a=A(0)/\Delta$, $b=B(0)/\Delta^2$, and $c=C(0)/\Delta^2$; while in the case $\eps = m \omega$, the definitions 
\begin{equation}
  d= B(0), \quad  f= - B(0)
\end{equation} 
have to be used.
\end{widetext}


\end{document}